# Systematic Review of Energy Efficient Thermal Comfort Control Techniques for Sustainable Buildings

Ghezlane Halhoul Merabet[1, 9,*] Mohamed Essaaidi[1], Mohamed Ben-Haddou[2], Basheer Qolomany[3], Junaid Qadir[4], Muhammad Anan[5], Ala Al-Fuqaha[6, 7], Mohamed Riduan Abid[8] and Driss Benhaddou[9]

[1] Smart Systems Laboratory (SSL), ENSIAS, Mohammed V University of Rabat, 713 Morocco
[2] MENTIS SA, 13, rue de Congrès, 1000 Brussels, Belgium
[3] Department of Cyber Systems, College of Business and Technology, University of Nebraska at Kearney (UNK), Kearney, NE 68849, USA
[4] Information Technology University, Lahore 54000, Pakistan
[5] Software Engineering Department, Alfaisal University-Riyadh, Saudi Arabia
[6] Information and Computing Technology (ICT) Division, College of Science and Engineering (CSE), Hamad Bin Khalifa University, Doha – Qatar
[7] Department of Computer Science, Western Michigan University, Kalamazoo, MI 49008, USA
[8] School of Science and Engineering, Alakhawayn University in Ifrane, 1005, Ifrane, Morocco
[9] Department of Engineering Technology, University of Houston, TX 77204, USA

**Abstract – Objective**. Different factors such as thermal comfort, humidity, air quality, and noise have significant combined effects on the acceptability and quality of the activities performed by the buildings' occupants who spend most of their times indoors. Among the factors cited, thermal comfort, which contributes to the human well-being because of its connection with the thermoregulation of the human body. Therefore, the creation of thermally comfortable and energy efficient environments is of great importance in the design of the buildings and hence the heating, ventilation and air-conditioning systems. In fact, among the strategies to improve thermal comfort while minimizing energy consumption is the use of control systems. Recent works have been directed towards more advanced control strategies, based mainly on artificial intelligence which has the ability to imitate human behavior. This systematic literature review aims to provide an overview of the intelligent control strategies inside building and to investigate their ability to balance thermal comfort and energy efficiency optimization in indoor environments. **Methods**. A systematic literature review examined the peer-reviewed research works using ACM Digital Library, Scopus, Google Scholar, IEEE Xplore (IEOL), Web of Science, and Science Direct (SDOL), besides other sources from manual search. With the following string terms: thermal comfort, comfort temperature, preferred temperature, intelligent control, advanced control, artificial intelligence, computational intelligence, building, indoors, and built environment. Inclusion criteria were: English, studies monitoring, mainly, human thermal comfort in buildings and energy efficiency simultaneously based on control strategies using the intelligent approaches. Preferred Reporting Items for Systematic Reviews and Meta-Analysis guidelines were used. **Results**. Initially, 1,077 articles were yielded, and 120 ultimately met inclusion criteria and were reviewed. **Conclusions**. From the systematic literature review, it was possible to identify the control methods used by the researchers, the most popular and efficient optimization strategies of thermal comfort and hence energy use in the built environments.

* Corresponding author: Ghezlane Halhoul Merabet (ghezlane.merabet@um5s.net.ma)

### Nomenclature

*Acronyms*

| | | | |
|---|---|---|---|
| ACMV | Air-Conditioning and Mechanical Ventilation | HMM | Hidden Markov Model |
| AI | Artificial Intelligence | HVAC | Heating, Ventilating, and Air Conditioning |
| AMV | Actual Mean Vote | IEA | International Energy Agency |
| ANN | Artificial Neural Networks | IEQ | Indoor Environmental Quality |
| APM | Advanced Predictive Methods | IPCC | International Panel of Climate Change |
| ARX | Autoregressive exogenous | KBS | Knowledge-Based System |
| ASHRAE | American Society of Heating, Refrigerating, and Air-Conditioning Engineers | KMA | K-Means Algorithm |
| BN | Bayesian Network | kNN | k-Nearest Neighbor |
| BPSOFMAM | Binary Particle Swarm Optimization Fuzzy Mamdani | LBMPC | Learning-Based Model Predictive Control |
| BPSOFSUG | Binary Particle Swarm Optimization Fuzzy Sugeno | MAS | Multi-Agent Systems |
| CA | Context-Awareness | MISO | Multi-Input, Single-Output |
| CI | Computational Intelligence | MOABC | Multi-Objective Artificial Bee Colony |
| CIBE | Chartered Institution of Building Services Engineers | MOPSO | Multi-Objective Particle Swarm Optimization |
| CTR | Comfort Time Ratio | MOGA | Multi-Objective Genetic Algorithm |
| DAI | Distributed Artificial intelligence | MPC | Model-based Predictive Control |
| DID | Degree of Individual Dissatisfaction | NIST | National Institute of Standards and Technology |
| DNN | Deep Neural Networks | OSHA | Occupational Safety and Health Administration |
| DRL | Deep Reinforcement Learning | PID | Proportional Integral Derivative |
| DT | Decision Tree | PMV | Predicted Mean Vote |
| DTR | Discomfort Time Ratio | PPD | Predicted Percentage of Dissatisfied |
| EACRA | Energy Aware Context Recognition Algorithm | PPV | Predicted Personal Vote |
| eJAL | Extended Joint Action Learning | RBF | Rule Base Function |
| EMS | Energy Management System | RL | Reinforcement Learning |
| FCM | Fuzzy Cognitive Maps | RNN | Recurrent Neural Networks |
| FLC | Fuzzy Logic Control | SCADA | Supervisory Control and Data Acquisition |
| FRB | Fuzzy Rule Base | TPI | Thermal Perception Index |
| GA | Genetic Algorithm | TSV | Thermal Sensation Vote |
| GDPR | General Data Protection Regulation | | |

*Symbols*

| | |
|---|---|
| $Temp_{set}$ | Set-point temperature |
| $N_i$ | The number of data points located within the boundary of the comfort zone for flat $i$ over one year |
| $n$ | The total number of flats in the building |
| $\xi$ | A weight factor $\in [0,1]$ |
| $\mu_T$ | The mean value of the optimal temperature set-point |
| $\sigma_T$ | The thermal comfort standard deviation |
| $t_n^{room}$ | The room temperature |
| $dev_n$ | The difference between room temperature and desired temperature $dt_n$ at any time $n$ |
| $M_t$ | The predicted thermal comfort value |
| $\Phi$ | The thermal comfort prediction algorithm |
| $T_t^{in}$ | The indoor air temperature at time slot $t$ |
| $H_t^{in}$ | The indoor humidity at time slot $t$ |
| $a, b, c$ | Constants defined in Kansas State University |
| $p_v$ | The vapor pressure |

## 1. Introduction & Background

Buildings (residential, commercial and industrial sectors) consume between 20% and 40% of the total final energy expenditure in developed countries, where half of this energy is consumed by the heating, ventilation, and air conditioning (HVAC) systems. This significant consumption has generated concern about the management and energy efficiency of buildings, from the economic, human behavior and scientific-technical point of views [1].

In this context, many factors influence the increase in energy consumption, and among them, the living standards of the population and the meteorological conditions. In fact, as more people spend significant time of their lives in artificially conditioned environments, thermal comfort has a direct effect on the productivity and satisfaction of each individual. For example, if the workplace environments do not provide the adequate thermal comfort conditions, the workers' performance decreases. However, most of comfort building management strategies tend to be not energy efficient . Hence there is a challenge in maintaining the thermal comfort of occupants while reducing energy expenditure. This challenge opens up great potential for research in the development of energy efficient control strategies that are designed to promote adequate thermal comfort.

Thermal comfort in built environments is a concept that has been challenging to model. In the last decades, a large number of indices have been established to analyze climates and HVAC control systems [2]–[5]. In common, there is the fact that measuring thermal comfort is not restricted to temperature measurement. Fanger [2], for example, proposed a method for estimating thermal comfort that, in addition to the temperature and the relative humidity, includes mean radiant temperature, air velocity and individual factors (such as metabolic rate and thermal resistance of clothing). Fanger proposed the predicted mean vote (PMV) as an index based on these variables. The closer to zero the value of the PMV, the better the thermal comfort sensation of occupants.

On the other hand, studies conducted on HVAC systems control used only temperature control, disregarding the influence of other parameters on thermal comfort. This is mainly due to the fact HVAC control system becomes expensive and complicated to manage by building owners if several input parameters are included. The typical HVAC control is performed through classic strategies such as ON/OFF control. Yet, the main disadvantage of these controllers is that they don't take into account the energy savings and poor regulation with temperature changes. Otherwise, the demands of better control require the use of sophisticated strategies that include another type of action mechanism, such as shading devices, automatic opening and closing of windows, and so forth [6], [7], with the main objective of minimizing the use of air conditioning systems and therefore energy consumption [8].

Nowadays, research works have been directed towards more advanced control structures that take multiple inputs (temperature, humidity, comfort perception, etc.) and uses, based on artificial intelligence or optimal intelligent control approaches based on expert systems, machine learning, deep learning, pattern recognition, neural networks, fuzzy systems, and evolutionary algorithms. Indeed, intelligent control systems have the ability to imitate human abilities, such as planning, learning, and adapting. The intelligent control field is multidisciplinary and combines techniques of artificial intelligence, control theory, heuristics, psychology, and operational research.

Generally, in the field of thermal comfort, there is no need to maintain the internal temperature and humidity at a fixed value; a range of values for these quantities creates a comfortable condition. From an economic point of view, it is interesting to find a good balance between thermal comfort and energy cost within the range of these values. This means that, reducing energy demand (hence cost) while maintaining thermal comfort indexes within a permissible range is a goal to be achieved in choosing appropriate control algorithms. For example, the fuzzy controllers have been used successfully in various applications related to thermal systems, as it is a suitable tool to imitate the behavior of building users, and developing linguistic descriptions of thermal comfort sensation, which approximate the PMV model calculations, and facilitate control systems. In this way, the fuzzy control scheme proposed by Oliveira et al. in [9] is characterized, by explicitly considering in the control theory, a range of admissible values for the internal temperature of the environment rather than a fixed value. In [10], Gouda et al. used a fuzzy PID control technique based on the use of a virtual PMV sensor. They compared two strategies of thermal comfort control in a building by considering the individual parameters (the metabolic rate and the clothes index) and the internal air velocity. In [11], the authors applied the fuzzy logic to control the climatic conditions (thermal, visual and air quality) inside a building. Another field of the application of fuzzy logic techniques is in solving optimization problems. For example, in [12] proposed a three-layer hierarchical control scheme based on fuzzy logic, with the main objective of authors of maximizing thermal comfort while optimizing energy efficiency inside a building. In this same line of work, authors in [13] proposed a thermal comfort control, using a Fuzzy PID controller, with automatic adaptation of proportional, integral and derivative parameters. In addition, the comfort control is done using the PMV index and constant individual variables.

Furthermore, other works explore the use of controllers based on the neural network, genetic algorithms or a combination of both methods for comfort control. In most cases, genetic algorithms have been introduced to solve optimization problems, since they are able to provide an optimal global solution. Regarding the genetic algorithm application, Yan et al. [14] describe how this technique can solve the problem of optimal control of the cooling source of an HVAC system, which has continuous and discrete control variables. An example of the application of the neural network was in [15], in which the authors proposed an approximation in order to determine the PMV index through neural networks for an HVAC system control. In [16] Liu et al. developed an internal thermal comfort assessment model based on the same technique by adjusting individual thermal comfort for each user. In this line of work, Lian and Du [17] presented a strategy of thermal comfort control, which uses a simplification of the PMV index, based on neural networks, able to acquire knowledge about the thermal comfort sensation of the occupants of an environment equipped with a climate control system.

Other related works on the application of advanced control systems in the field of comfort and energy saving in buildings used other techniques. For instance, in [18] Hadjiski et al. developed a supervisory system consisting of a predictive controller applied to an HVAC system in order to keep the PMV index as close to zero as possible, and to minimize the energy consumption in the cost functions of model-based predictive controllers (MBPC). In [19] Donaisky et al. proposed two predictive control structures based on a restricted model for the thermal comfort control inside a building. This strategy takes into account climate predictions to increase energy efficiency while satisfying the comfort conditions for users.

Due to the need to investigate the appropriate method for intelligent control systems for energy and thermal comfort of buildings, as well as identifying recommendations for future research, this review reviews and maps all existing studies and builds a classification scheme that will help to elucidate AI/ML-based models used for further studies and evaluations. This article aims to obtain a holistic view of the challenges of providing thermal comfort to the users inside buildings in an energy efficient way, and to produce bibliographic material to help researchers and professionals in the area to undertake such a challenge.

## 2. Systematic Literature Review Methodology

The literature review process followed the criteria established by the PRISMA guidelines, designed to guide studies of systematic review and review by meta-analysis [20], [21]. We conducted a literature search using ACM Digital Library (https://dl.acm.org/), Scopus (https://www.scopus.com/home.uri), Google Scholar (https://scholar.google.gr), IEEE Xplore (IEOL) (https://ieeexplore.ieee.org/Xplore/home.jsp), Web of Science (https://mjl.clarivate.com/home), and

Science Direct (SDOL) (https://www.sciencedirect.com) to identify the peer-reviewed publications related to the intelligent comfort control. The mentioned databases are chosen for being repositories of the main scientific publications of impact and relevance for the analyzed area. In addition to the search in the references of the selected articles themselves and through manual research. Figure 1 shows the research steps adopted in this review.

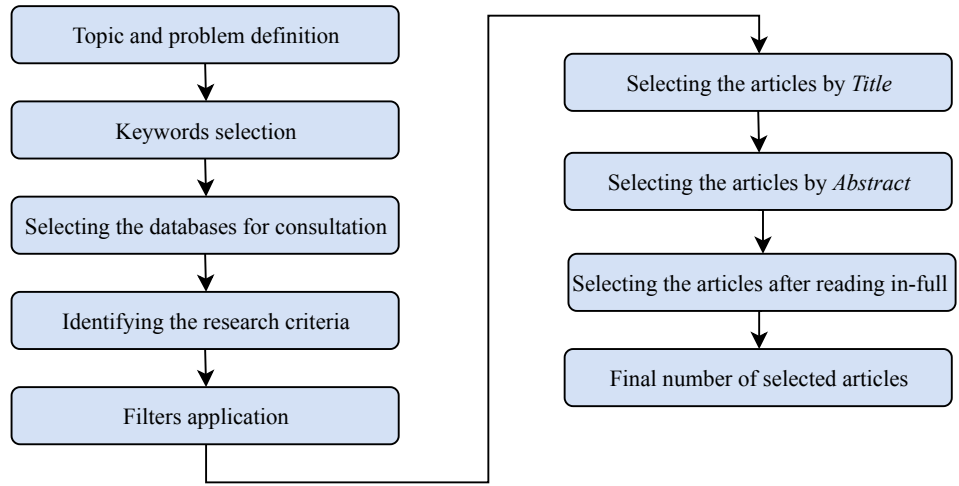

**Figure 1.** Stepwise structure of search process.

For the review, we selected and analyzed original peer-reviewed articles and journal publications published within a period from 1992 to 2020, and whose main topic is studying the development of artificial intelligence-based techniques used to develop control systems reducing energy consumption while maintaining thermal comfort of occupants inside buildings. After applying filters according to the search protocol illustrated in Figure 2, a total of 120 results articles have fitted the inclusion criteria: (1) studies performed in indoor environments; (2) works presenting innovative AI-based tools and their deployment in HVAC and thermal comfort control; and (3) describing the system performance after applying the AI control tools.

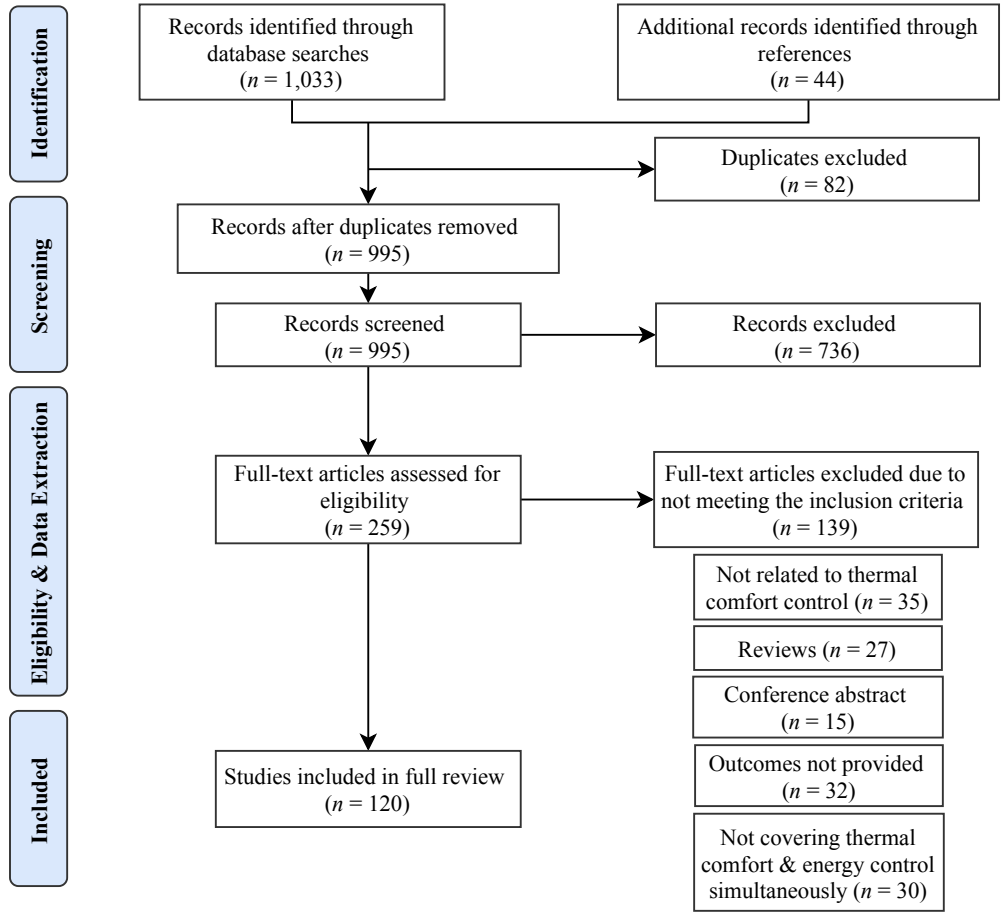

**Figure 2.** Flowchart of the articles' selection process.

## 3. Review Results

The systematic extraction of information allowed the elaboration of Table 1, which summarizes the 120 articles and presents the results classified according to the (1) year of publication, (2) study/or academic case, (3) source of data used for models' evaluations, (4) AI and ML assisted tool, (5) application scenarios, (6) models and methods used to measure the thermal comfort, (7) input(s) and control parameter(s) as well as (8) key results and findings.

**Table 1.** Descriptive analysis of the adherent works for the review using AI-assisted techniques for energy and thermal comfort management in buildings.

| Year | Study Case | Source of Data used | Underlying AI/ML techniques | Application Scenario | Thermal comfort-based model/method | Input(s) & Controlled parameter(s) | Key Results | Ref |
|---|---|---|---|---|---|---|---|---|
| 1993 | An intelligent operation support system (IOSS) to improve HVAC operations for IAQ control and energy saving for industrial application. | Field survey conducted by the authors (in an office building at the University of Alberta, Canada) | Knowledge-based system (KBS) development to optimize and automate the HVAC process | Optimized setting | PMV (Fanger's method) | HVAC system structure (conditioning type, zone type, air volume, dehumidifying unit); Weather; Indoor setting; Supply air parameters; Clothing; Activity level | Based on the integrated distributed intelligent framework, the developed system can provide real-time planning, and assisting the interaction between the operator and the HVAC process | [22] |
| 1998 | Fuzzy controller development for improving thermal comfort and energy saving of HVAC systems. | Interviews | Fuzzy logic controller development | Fuzzy control to enhance control performance | PMV (Fanger's model) | Air temperature; Relative humidity; Air velocity; Mean radiant temperature; Occupants' activity level; Clothing insulation | The proposed system allows the user to compromise solution (comfort requirements /energy saving) | [23] |
| 1999 | Multi-objective supervisory control of building climate and energy. | Interviews | Fuzzy-based supervisory controller development | Optimized setting | Pre-defined (standardized) indoor temperature (i.e., between 20ºC and 24ºC) | Outside/inside temperature; Outside/inside relative Humidity; Outside/inside $CO_2$ concentration; Occupancy; Weighting comfort/economy factor | The proposed system allows the user to compromise solution (comfort requirements /energy saving) | [24] |
| 2001 | PMV-based fuzzy logic controller for energy conservation and indoor thermal requirements and of a heating system in a building space. | Weather data collection/Interviews | Fuzzy logic controller development | Fuzzy control for thermal sensation investigation to improve control performance | PMV (Fanger's model) | PMV and PPD calculated from: Internal air temperature; Mean radiant temperature; Relative humidity of the internal air (activity level and clothing considered constant) | While maintaining PMV index at 0 and PPD with a maximum threshold of 5%, fuzzy controller had better performance with a heating energy of 20% (compared with conventional tuned PID control). | [10] |
| 2001 | Developing fuzzy controller for energy saving and occupants' thermal-visual comfort and IAQ requirements. | Interviews/Indoor climatic data | Three fuzzy controllers including fuzzy PID, fuzzy PD and adaptive fuzzy PD | Fuzzy control to enhance control performance | PMV (Fanger's model) | PMV index; Users preferences; $CO_2$ concentration; Illuminance level | Adaptive fuzzy PD showed the best performance for energy consumption which can reach up to 25-30% and the PMV/CO2 responses, while for visual comfort, the non-adaptive fuzzy PD was sufficient. | [25] |
| 2002 | Controller development for indoor environmental conditions management for users' satisfaction while minimizing energy consumption inside a building. | Indoor/Outdoor climatic (instrumental) data | GA-based fuzzy controller development | Optimized setting (through GA) | PMV (Fanger's model) | PMV index; User's preferences; Indoor/Outdoor temperature; $CO_2$ concentration; The rate of change of $CO_2$ concentration; Indoor illuminance; Indoor air velocity; Indoor humidity | Overall energy saving up to 35%, with a steady-state error of 0.5 for PMV, ~80ppm for CO2, and ~80 lx for illuminance (after applying GA). | [26] |
| 2003 | Developing controller for HVAC system to improve indoor comfort requirements and energy performance in two real sites. | Indoor/Outdoor climatic data | GA-based fuzzy controller development | Optimized setting (through GA) | PMV (Fanger's model) | PMV index; User's preferences; Indoor/ Outdoor temperature; $CO_2$ concentration; The rate of change of $CO_2$ concentration; Indoor illuminance; Indoor air velocity; Indoor humidity | While maintaining a steady-state indoor conditions, the developed controller showed best experimentation results in the real test cells, with up to 30% energy saving for CNRS–ENTPE case and 12.5% for ATC. | [27] |
| 2003 | Fuzzy controller for indoor environment management. | Indoor/Outdoor climatic data/questionnaires | Five fuzzy controllers including: fuzzy P, fuzzy PID, fuzzy PI, and adaptive fuzzy PD | Fuzzy logic control to improve control performance | PMV (Fanger's model) | PMV index; Outdoor temperature; $CO_2$ concentration; Indoor illuminance; The rate of change of $CO_2$ concentration | By maintaining PMV index between 0 and 0.1, and $CO_2$ concentration less than 20 ppm; the fuzzy P-controller showed the best performance, with heating and cooling energy saving up to 20.1% | [28] |
| 2003 | Fuzzy control for indoor environmental quality, energy and cost efficiencies | Weather data (Kew, UK)/Indoor climatic data/Interviews | Fuzzy logic controller development | Fuzzy logic for control decision | Defined ranges/ Preferred set-points variables | Zone (indoor) temperature; $CO_2$ concentration; Relative humidity | Fuzzy approach showed its ability to deal with multivariate problems by collaborating expert knowledge for decision making at complex level, with no significant differences with conventional controls in energy and cost efficiencies and IAQ performance. | [29] |

**Table 1.** Cont.

| Year | Study Case | Source of Data used | Underlying AI/ML techniques | Application Scenario | Thermal comfort-based model/method | Input(s) & Controlled parameter(s) | Key Results | Ref |
|---|---|---|---|---|---|---|---|---|
| 2004 | Two-objective optimization of HVAC system control with two variable air volume (VAV) systems | Recorded data of a VAV system/Indoor climatic data | GA-based supervisory control | Optimized setting to determine the optimal set-points of control | PMV-PPD (Fanger's model) | Zone temperature set-points; Supply duct static pressure set-point; Supply air temperature set-point | The on-line implementation of GA optimization allowed to save up to 19.5% of energy consumption while minimizing the zone airflow rates and satisfying thermal comfort | [30] |
| 2005 | Development of fuzzy rule-based controller using GA for HVAC system | Data collection using a real test-site/Manufactured data/Interviews | Fuzzy logic controller development and optimized via GA | Optimized setting using GA for rule weight derivation/selection performance | PMV (Fanger's model) | PMV index; Difference between supply and room temperatures; $CO_2$ concentration; Outdoor temperature; HVAC system actuators | By considering the rule weights and rule selection, results showed that FLC controller presented improvement by 14% in energy saving and about 16.5% in system stability | [31] |
| 2005 | Controller development to improve energy conservation with a constraint on the individual dissatisfactions of indoor environment | Meteorological year weather files of 3 different cities/Interviews (50 random population)/Indoor climatic data | Fuzzy logic control based on Nearest Neighbors (kNN) approximations | Gradient-based optimization | $DID(vote) = \frac{1 + \tanh(2|vote| - 3)}{2}$ | Neighboring office temperatures; Desired temperatures of each individual | While maintaining the population dissatisfaction under 10%, experimental results showed that the Optimized HIYW presented better performance than OFSA (PPD exceeding 20% for ~15% of population and 50% for ~5%) to minimize the energy consumption | [32] |
| 2005 | Decentralized system development for controlling and monitoring an office building | Field experiments at test-site in a real building (Villa-Wega): Climatic data/Interviews (GUI via PDA) | Agent-based approach deployment for energy use control and customer satisfaction | Distributed AI | Personal comfort based on individual preferences for each user | Occupant preferences; Room occupancy; Temperature; Light intensity | The MAS approach allowed to save up to 40% of energy, compared to thermostat approach and almost 12% compared to timer-based approach. The reactive approach is slightly more energy consuming than the pro-active, which ensures 100% of thermal satisfaction to the users | [33] |
| 2005 | NN-based control development for individual thermal comfort optimization, and energy saving by combining a thermal space model for VAV HVAC application. | N/A (Simulation data) | Direct neural network development for better performance of comfort control | Optimized setting | PMV (Fanger's model) | Activity level; Clo value; Indoor air temperature; Mean radiant temperature; Relative air velocity; Humidity Dimension of thermal space; Cooling/heating load; HVAC capacity; Air flow rate; Mixed air ratio; Outdoor temperature/Humidity ranges | Compared to the conventional HVAC systems, the proposed controller showed high comfort level (by maintaining the comfort zone between -0.5 and +0.5) while conserving energy. However, there still some limitations in practice. | [17] |
| 2005 | Integrated indoor environment energy management system (IEEMS) implementation for buildings application | Experiments conducted in real office buildings: Climatic data/ Subjective data using Kiosk smart card [34] | Fuzzy controller development for users' comfort fulfillment | Fuzzy indoor comfort controller | PMV (Fanger's model) | PMV index; Outdoor temperature; $CO_2$ concentration; The rate of $CO_2$ concentration; Indoor illuminance | Up to 38% energy conservation in both buildings without compromising the indoor comfort requirements. | [35] |
| 2005 | Dynamic illumination and temperature response control in real time conditions | Test chamber built in a Faculty of Civil Engineering in Slovenia: Climatic data | Fuzzy + proportional-integral-differential (PID) controller development for improving comfort control performance | Fuzzy logic for indoor thermal and visual comfort optimization | Temperature preference set-point (by the user) | Inside/Outside temperature; Solar direct/Reflected radiation; Inside illumination; Current roller blind position | Adjusting automatically roller blind position and window geometry according to external weather enables to get closer to thermal-visual preferences, which contributes to lower energy consumption for lighting, heating and cooling and cost-saving enhancement | [36] |
| 2006 | Centralized HVAC with multi-agent structure | Experiments/Climatic data | Multi agent-based structure development for thermal comfort control + optimized setting via ACO | Distributed AI and optimized setting based ACO | PMV (Fanger's model) | Air temperature; Radiant temperature; Relative humidity; Air velocity | The control accuracy goes around 89% to 92.5%. which means that the thermal comfort is predicted by 7.5% to 11% of error rate | [37] |

| Year | Study Case | Source of Data used | Underlying AI/ML techniques | Application Scenario | Thermal comfort-based model/ method | Input(s) & Controlled parameter(s) | Key Results | Ref |
|---|---|---|---|---|---|---|---|---|
| 2006 | Adaptive fuzzy control strategy for comfort air-conditioning (CAC) system performance | Experiments conducted in an artificial environment chamber in office buildings in China: Climatic data | Indirect adaptive fuzzy model control strategy applied to improve thermal comfort and energy saving | Fuzzy adaptive controller development | PMV (Fanger's model) | PMV index calculation by measuring indoor temperature and assuming the five variables affecting thermal comfort as constant | The adaptive fuzzy controller could save almost 18.9% of energy, compared to PID controller. Results showed that the fuzzy controller has given rise to much more comfortable thermal conditions | [38] |
| 2007 | Linear reinforcement learning controller (LRLC) for energy saving while sustaining comfort requirements. | Testing environment: Climatic data/Interviews | Linear reinforcement learning controller development instead of using traditional on/off controller | Machine learning and adaptive user satisfaction simulator | PMV-PPD (Fanger's model) | Indoor/outdoor temperature; Relative humidity; $CO_2$ concentration | Over a period of 4 years, training the LRLC, the energy consumption has been increased from 4.77Mwh to 4.85Mwh, however the PPD index has been decreased from 13.4% to 12.1% | [39] |
| 2007 | Development of an intelligent coordinator of fuzzy controller-agents (FCA) for indoor environmental control conditions using 3-D fuzzy comfort model | Weather data/Simulation data (Climatic/ Subjective data) | PI-like FLC standing of FCA with intelligent coordination | Intelligent control system-based fuzzy logic approach | PMV (Fanger's model) | PMV index; Illuminance level; $CO_2$ concentration | The fuzzy controller-agent (FCA) with the intelligent coordinator (IC) showed significant results by maintaining the controlled variables in acceptable ranges (PMV between -0.5 and +0.6) besides up to 30% of energy savings | [11] |
| 2007 | Modelling indoor temperature using autoregressive models for intelligent building application | Surveys and field experiments in 26 air-conditioned and 10 naturally ventilated classrooms (Indoor/Outdoor climatic data) | Determining the adequate structure of autoregressive model with external input (ARX) and autoregressive moving average model with external input (ARMAX) for indoor temperature prediction | Predictive control | Black-box model to predict indoor temperature based on variables: $T_o$, $R_a$, $V_w$, $R_{HO}$ | Outside air temperature ($T_o$); Global solar radiation flux ($R_a$); Wind speed ($V_w$); Outside air relative humidity ($R_{HO}$) | While continuously monitoring the energy consumption to enable energy savings. Results showed that ARX model gave better temperature prediction than ARMAX model by the structure ARX(2,3,0) with a coefficient of determination of 0.9457 and the ARX(3,2,1) with a coefficient of determination of 0.9096. | [40] |
| 2007 | Fuzzy controller development for improving the indoor environmental conditions while reducing energy requirements for building energy management system | Indoor climatic data/Interviews | Fuzzy control techniques + Man Machine Interface to satisfy the users preferences | fuzzy control for improving control performance | PMV (Fanger's model) | PMV index; Illuminance level; $CO_2$ concentration; Users preferences | Using a suitable cost function for BEMS allowed to save energy at a level lower than recommended by the literature. Also, the users were satisfied by the adoption of the fuzzy controller | [41] |
| 2008 | Intelligent comfort control system (ICCS) design by combining the human learning and minimum energy consumption strategies for HVAC system application | Interviews/Climatic data | DNN-based controller to maintain PMV within the comfort zone for better control performance deep | Optimized setting | PMV (Fanger's model) | Comfort level (PMV index); Air temperature; Relative humidity; Air velocity; Fresh air flow rate; System to fresh air flow rate ratio; Air change | By applying the VAV control, the system could save energy whilst a higher comfort level was satisfied compared with the conventional temperature controller by maintaining the PMV within the comfort zone. | [42] |
| 2009 | Developing an inferential sensor based on the adaptive neuro-fuzzy modeling to estimate the average temperature in space heating systems | Experimental data from a laboratory heating in Italy (Climatic data) | ANFIS development to improve the heating systems performance | Fuzzy logic and adaptive neuro fuzzy inference system (ANFIS) | Estimating the average air temperature based on $T_o$, $Q_{SQL}$, and Fire | External temperature ($T_o$); Solar radiation ($Q_{SQL}$); boiler control signal (Fire) | The average air temperature estimated by ANFIS control model are very close to experimental results, with a highest possible RMSE = 0.5782°C. | [43] |
| 2009 | Exploring the impact of optimal control strategies of a multi-zone HVAC system on the energy consumption while maintaining thermal comfort and IAQ of a built environment. | Experiment conducted in an academic building in Lebanon (Climatic data) | GA development for the optimization of HVAC control | Optimized setting and predictive control | PMV (Fanger's model) | Supply temperature; Fresh air amount; Supply flow rate, PMV index | Up to 30.4% savings in energy costs when compared to conventional base strategy whilst sustaining comfort and indoor air quality | [44] |

**Table 1.** Cont.

| Year | Study Case | Source of Data used | Underlying AI/ML techniques | Application Scenario | Thermal comfort-based model/method | Input(s) & Controlled parameter(s) | Key Results | Ref |
|---|---|---|---|---|---|---|---|---|
| 2009 | Predicting fan speed based on ANFIS for energy saving purpose in HVAC system | Experimental study (Simulation data/Climatic data) | PID + ANFIS model to predict fan motor speed in HVAC system | Predictive control | Desired temperature by controlling the damper | Ambient temperature; Fan motor speed; Damper gap rates | Simulation results showed that the ANFIS model is more effective and can be used as an alternative for HVAC control system. Statistically, RMS and R-squared were used for model validation in different zones (Zone-1: RMS = 15.6750 and R2 = 0.9402; Zone-2: RMS = 17.7019 and R2 = 0.9410; Evaporator: RMS = 3.3475 and R2 = 0.9954). | [45] |
| 2009 | Estimating occupant mental performance and energy consumption of determining acceptable thermal conditions under different scenarios | Observations recorded from field studies (in 2 real buildings in Singapore and Sydney)/Data-base RP-884 [46] | Bayesian Network (BN) model was used to infer the probability of the occupants' thermal satisfaction | Predictive control | PMV (Fanger's model) and the adaptive comfort model | Building configuration; Operative temperature; Clothing insulation; Outdoor temperature | Two building configurations (with/without mechanical cooling) were used for simulation under different climate regions (tropical/subtropical). Results concluded that determining acceptable thermal conditions with the adaptive model of comfort can result in significant energy saving with no large consequences for the mental performance of occupants. | [47] |
| 2010 | Energy consumption optimization and thermal comfort management using data mining approach in built environment | Climatic data collected from a test bed installed in an academic building (ERI) in Ireland | Decision tree classifier (C4.5 algorithm) model was used to predict thermal comfort under different environmental conditions | Predictive control and optimized setting | Comfort levels based on CIBSE standard: comfort temperature in offices is between 21ºC and 23ºC | $CO_2$ level; Humidity; Outside air humidity; Outside wind speed; Under floor input flow temperature; Under floor output flow temperature | Based on decision tree analysis and results relying ambient environmental conditions with user comfort, designers and facility managers can determine the optimal energy use. | [48] |
| 2010 | Multi-objective optimization methodology used to optimize thermal comfort and energy consumption in a residential building | Data collected from numerical experiments conducted, using 100 population, in residential buildings in Canada | - ANN used in simulation to characterize the building behavior<br>- ANN and NSGA-II were combined for optimization and fast evaluations | Optimized setting | PMV (Fanger's model) | **HVAC-related variables** (Heating/cooling temperature; Relative humidity; Supply air flowrates; Thermostat delays) and **Building-related variables** (Thermal mass; Window sizes) | Optimization results showed considerable improvement in thermal comfort (average PMV<4%), and reduction in energy consumption (relative error< 1%) for the total energy consumption. Simulation time was reduced compared to the classical optimization methods. | [49] |
| 2011 | Intelligent control system development to optimize comfort and energy savings using soft computing techniques for building application | Simulation data (TRNSYS 16 and MATLAB software) | PI-Like fuzzy logic controller optimization with GA | Optimized setting | PMV (Fanger's model) | PMV index; Illumination level; $CO_2$ concentration | While maintaining the PPD index within acceptable limits, i.e., below 10%, the proposed system has successfully managed the user's preferences for comfort requirements and energy consumption. | [50] |
| 2011 | Controller development for a heating and cooling energy system | Simulation data | GA-based fuzzy PID (GA-F-PID) controller development | Predictive control | Fixed set-point temperature for the thermal zone (24ºC) | Temperature | The proposed methodologies allowed to achieve higher energy efficiency and comfort requirements by lowering equipment initial and operating costs up to 35%, and comfort costs up to 45%. | [51] |
| 2011 | Multi-agent simulation for building system energy and occupants' comfort optimization | Simulation data collected from a test bed in a commercial facility building (including students, faculty and stuffs occupants) in Los Angeles, CA. | Multi-agent comfort and energy simulation (MACES) implementation + Proactive-MDP optimization for building and occupants' control and management | Distributed AI | PMV (Fanger's model) | PMV index; Building location; Outdoor temperature; Real-time occupancy; Time of day; Glazing areas | 17% energy savings while maintaining high comfort level, approximately 85% occupants' satisfaction. | [52] |

| Year | Study Case | Source of Data used | Underlying AI/ML techniques | Application Scenario | Thermal comfort-based model/method | Input(s) & Controlled parameter(s) | Key Results | Ref |
|---|---|---|---|---|---|---|---|---|
| 2011 | AI-based thermal control of a typical residential building in USA | Simulation data based on the American Housing Survey for 2 residential buildings in USA [53] | ANFIS development and control performance comparison with ANN | Fuzzy logic and adaptive neuro fuzzy inference system (ANFIS) | Defined comfort bands (20–23°C for winter and 23–26°C for summer) from temperature set-points (21.5°C winter and 24.5°C summer) | Dry-bulb/Wet-bulb temperature of the air stream entering the condenser; Mass flow rates of air/water/ refrigerant/ pressure/ temperature of the refrigerant | ANFIS-based control method could save 0.3% more energy than the ANN in the winter/summer periods, ANFIS could save 0.7% more energy. Both methods satisfied thermal comfort periods (~98% in the winter and 100% in the summer), with reduced standard deviations of air temperature from the set-point for both seasons (under 0.2°C). | [54] |
| 2011 | Fuzzy adaptive comfort temperature (FACT) model development for intelligent control of smart building. | Interviews/Daily average temperature data collected from [55], of an area around Toledo in USA | FACT + Grey prediction for multi-agent control system + optimized setting through PSO | Fuzzy with grey prediction control and optimized setting through PSO | Adaptive comfort model | Customer's preference; Outdoor environmental information (average daily temperature); Online energy production information | Using the FACT model with grey predictor in agent-based control system of a smart building, provided reasonable comfort temperature with less energy consumption to the customers | [56] |
| 2011 | Developing a MAS combined with an intelligent optimizer for intelligent building control | Indoor climatic data/Interviews | Coupling MAS and PSO to improve the intelligence of a multi-zone building | Optimized setting | Temperature set-point control | Occupant's preference; Temperature; Illumination level; $CO_2$ concentration | The implementation of PSO optimizer allowed to maintain a high-level of overall comfort, i.e., mainly around 1.0, when the total energy supply was in shortage. | [57] |
| 2012 | Improving the energy efficiency in an AC by reducing transient and steady-state electricity consumption on BRITE (Berkeley Retrofitted and Inexpensive HVAC Testbed for Energy Efficiency) platform. | Experimental data measurements using BRITE testbed | Learning-based model-predictive control (MPC) development for maintaining comfort temperature | Learning-based model predictive control | Comfort specifications based on OSHA guidelines (20°C – 24.2°C) | Occupancy; Temperature | 30%–70% reduction in energy consumption while maintaining comfortable room temperature by keeping temperature close to the specified comfort middle (22°C) | [58] |
| 2012 | Model-based predictive control development for thermal comfort improvement with auction of available energy of a limited shared energy resource in three houses. | Simulation data | Distributed model predictive control (DMPC) to obtain comfortable indoor temperature by considering the available energy limitations | Distributed model predictive control | Defined comfort temperature bounds (i.e., comfort zones) | Indoor temperature; Occupancy; Building thermal characteristics; Load disturbances profile | The developed system is flexible, in a way allowing the customer to shift between comfort and lower cost. By knowing disturbances profile, agents can make their bid to get significant savings. | [59] |
| 2012 | A discrete model-based predictive control for thermal comfort and energy conservation in a building of the University of Algarve | Weather data collection/ Experiment data measurements conducted in an office building | Radial basic function (RBF) ANN development to estimate comfort level (PMV) + MOGA used with MBPC for models' selection | Discrete models-based predictive control | PMV (Fanger's model) | Outdoor air temperature; Outdoor air humidity; Global solar radiation; Indoor air temperature; Indoor air humidity; Globe temperature; Windows/Doors state; Activity | Up to 50% energy savings are achieved by using the MBPC, which provided good coverage of the thermal sensation scale, when used with radial basis function-NN models. | [60] |
| 2012 | Coordinating occupants' behaviors for thermal comfort improvement and energy conservation of an HVAC system | Actual building and occupants' data measurements from a real-world testbed implemented in a university building in LA, USA | Distributed AI development to achieve multi-agent thermal comfort and building energy control | Distributed AI | PMV (Fanger's model) | Real-world feedback data; Building/occupant data; Occupant suggestions | Reducing 12% of energy consumption while maintaining 70%–75% occupant satisfaction for both proactive and proactive-MDP (showed by the distributed evaluation) | [61] |
| 2012 | Distributed AI control with information fusion-based Indoor energy and comfort management for smart building application | Simulation data | Multi-agent control system with heuristic optimization (PSO) development to enhance the comfort level and reduce energy consumption | Distributed AI | Setting comfort range as: $[T_{min}, T_{max}] = [19.4°C, 24.4°C]$ | Customer's preference; Illumination level; $CO_2$ concentration; Air temperature | All case studies showed the effectiveness of the system of the developed system in different operating scenarios | [62] |

| Year | Study Case | Source of Data used | Underlying AI/ML techniques | Application Scenario | Thermal comfort-based model/method | Input(s) & Controlled parameter(s) | Key Results | Ref |
|---|---|---|---|---|---|---|---|---|
| 2013 | Model-based predictive control development for optimal personalized comfort and energy consumption management in an office workplace | Occupancy and temperature data collection using SPOT+ system within a workspace environment | k-nearest neighbor (kNN) algorithm was used for occupancy prediction + LBMPC-based model for temperature prediction | Predictive control and optimized setting | PPV function defined as an affine transform of $pmv$: $ppv = f_{ppv}(pmv)$ | Indoor temperature; Occupancy | While maintaining personal comfort (PPV) in a range of $[-0.5, +0.5]$, about 60% energy savings when compared with fixed temperature set-point, and discomfort reduction from 0.36 to 0.02 compared to baseline methods. | [63] |
| 2013 | Fuzzy method-based data-driven to model and optimize thermal conditions of smart buildings applications. | Thermal comfort survey (online questionnaire) | Type-2 fuzzy sets based for modeling thermal comfort words and energy consumption | Fuzzy control and optimized setting | Comfort temperature ranges defined by the users | Air temperature | The type-2 fuzzy model performs better, with $RMSE = 12.55$ compared to the linear regression model where the $RMSE = 17.64$. Also, the multi-objective optimization could recommend reasonable temperature interval giving comfortable sensations while reducing energy consumption. | [64] |
| 2013 | Intelligent control system deployment for energy and comfort management in commercial buildings | Simulation data | MAS development for energy and comfort management + fuzzy logic control (FLC) with optimized setting | Distributed AI and fuzzy logic control (FLC) with optimized setting | Comfort temperatures according to set values by the users (preferences) | Customer's preference; Illumination level; Indoor temperature | Up to 0.9 is achieved by comfort factors, i.e., the customers satisfaction is ensured. The GA-based optimization allowed to minimize the energy consumption | [65] |
| 2013 | Identifying building behaviors related to energy efficiency and comfort for an office building in the Pacific Northwest | Data measurement using sensors throughout a test building. | Implementing a fuzzy knowledge-base for building behavior extraction. | Fuzzy rule base and optimized setting | Comfort levels based on average zone temperature | Time; Outside air temperature; Chiller temperature; Mixed air temperature; Return air temperature; Damper position; Exhaust fan load/current; supply fan load/current; Zone temperature | The developed framework was able to identify and extract complex building behavior, which improve the building energy management systems (BEMSs) by eliminating the low efficiency and low comfort behavior | [66] |
| 2013 | Intelligent management system development for energy efficient and comfort in building environments | Distributed sensors to collect: Environmental data, Occupancy data and Energy data | User-centered control based on behavior prediction | Distributed AI | Individual thermal comfort based on the indoor temperature | Indoor/Outdoor temperature; Illumination level; $CO_2$ concentration; Users' preferences | Indoor thermal comfort is considered to be highly satisfactory to users while maintaining a comfort level around 0.61 (PMV). Case studies simulation results showed that the developed MAS could manage comfort needs and reduce energy consumption simultaneously | [67] |
| 2014 | Dynamic and automatic fuzzy controller for indoor for indoor thermal comfort requirements | Recorded data using a real testbed scenario | ANN with NNARX-type performs the weather forecasting to feed a fuzzy logic controller | Predictive control | Building comfort scale (temperatures ranges) based on personal comfort preferences | Dry bulb outdoor/indoor air temperature; Relative humidity; Wind speed | The proposed control system allowed to achieve efficient use of energy and bring the room temperature to the maximum value of personal comfort. | [68] |
| 2014 | Predicting an integrated building heating and cooling control based on weather forecasting and occupancy behavior detection in the Solar House test-bed in real-time located in Pittsburgh | Manufacture datasets/Data (weather and occupancy) measurements through a real-time experiment | GMM + HMM were used for occupancy behavior model development; HM + AGP were implemented for weather forecasting; and a Nonlinear model predictive control (NMPC) was designed for heating/cooling system | Predictive control and optimized setting | Learning personal comfort temperature set-points for cooling/heating seasons based on the weather and occupancy information | Indoor temperatures; Indoor relative humidity; $CO_2$ concentration; Lighting; Motion; Acoustics; Power consumption for electrical plugs/HVAC/ lighting systems | 30.1% of energy reduction in the heating season, besides 17.8% in the cooling season when comparing to the conventional scheduled temperature set-points. Also, the use of NMPC allowed reducing time not met comfort (from 4.8% to 1.2% in heating season, and from 2.5% to 1.2% for cooling season). | [69] |
| 2014 | Reinforcement learning for tenant comfort and energy use optimization in HVAC systems | Simulation data | RL (Q-learning)-based supervisory control approach | Optimized setting | Occupant's comfort is achieved by learning from the tenant preferences and occupancy patterns | Time; Tenant thermal preferences; Temperature; HVAC state; Occupancy patterns | Learning to adjust/schedule, appropriately, thermostat temperature setpoints for energy efficiency while keeping the tenant comfortable | [70] |

**Table 1.** Cont.

| Year | Study Case | Source of Data used | Underlying AI/ML techniques | Application Scenario | Thermal comfort-based model/ method | Input(s) & Controlled parameter(s) | Key Results | Ref |
|---|---|---|---|---|---|---|---|---|
| 2014 | Improving HVAC systems operations by coupling personalized thermal comfort and zone level energy consumption for selecting energy-aware and comfort-driven set-points | Questionnaire (subjective comfort data)/Experiment data collection | Knowledge-based approach to optimize the air flow rates performance for HVAC system | Optimized setting | Personalized comfort profiles based on the participatory sensing approach by adopting TPI scale showing thermal votes ranging from -5 to +5 | Personal comfort data (Thermal preferences index); Room temperature profile (set-point, outside temperature, occupancy); Energy data (airflow profile) | About 12.08% (57.6m3/h) average daily air-flow rates were reduced in three target zones, compared to operational strategy that focus on comfort only. | [71] |
| 2014 | Improving the fuzzy controller's performance for comfort energy saving in HVAC system | Simulations (BPS tool EnergyPlus, SketchUp, MATLAB and BCVTB) using: Weather data (Toronto, Canada)/Real building model data (Hotel located in Toronto, Canada) | GA-based tuning for FLC optimization | Fuzzy control and optimized setting | Individual comfort classes: ISO 7730 based on PMV/PPD (Fanger's model) | **Environmental parameters:** Ambient air temperature; Mean radiant temperature; Relative humidity; Relative air velocity; Clothing insulation; Metabolic rate. **Building parameters:** Exterior walls; Exterior windows; Exterior door; Exterior floor; Exterior roof and Interior celling; Interior doors | While maintaining the GAFLC operations with PMV limits of $|PMV| \leq 0.7$. The overall energy consumption is decreased by 16.1% in case of cooling and 18.1% in case of heating. Also, the PMV is reduced from -0.3735 to -0.3075 compared to EnergyPlus. | [72] |
| 2014 | Radiator-based heating system optimization to maintain indoor thermal comfort and minimize the energy consumption for residential building | Simulations (MATLAB) using a real building model data/Indoor climatic and occupancy data measurements/Weather data | Three optimization algorithms were tested: GA, PSO and SQP to be combined with random neural network (RNN) control model to calculate the optimal control input | Predictive control and optimized setting | PMV-based set-point, defined by the Institute for Environmental Research at KSU under ASHRAE contract $PMV = a \cdot t + b \cdot p_v - c$ | Current room air temperature; Outside temperature; Number of occupants; Flowrate of inlet water for radiator | The proposed model accuracy is of MSE=38.87% for PSO less than GA/ MSE=21.19% for PSO less than SQP. RNN with GA allowed to maintain comfortable comfort conditions with the minimum energy consumption (400.6 MWH), compared to MPC model. | [73] |
| 2014 | Deploying and evaluating a user-led thermal comfort driven HVAC control framework in office building on University of Southern California | Field study data collection: Questionnaire (subjective comfort data)/Outdoor and Indoor environmental data. | Fuzzy predictive model used to learn the user's comfort profiles. | Predictive control | Personalized comfort profiles based on TPI scale | User's preference vote; Temperature; Humidity; $CO_2$ concentration; Light intensity | The developed framework showed promising results for energy saving and comfort improvement. 39% reduction in daily average airflow rates (when HVAC conditions at user's desired temperature). | [74] |
| 2014 | A human and building interaction toolkit (HABIT) development for building performance simulation | Field data on comfort and behavior from a real air-conditioned office building in Philadelphia. | Coupling Agent-based model (ABM) and adaptive behaviors for energy use and thermal comfort management | Distributed AI | Sensation and acceptability ranges modeled via individual distributions, based on PMV index (Fanger's model) | Occupants behaviors profiles (Clothing adjust, Fan On, Heater On, Thermostat Up/Down, Window open); Indoor operative temperature | Up to 32% reduction of total energy use in all zones in summer without significant increase in winter are expected if building managers embrace the use of lower energy local heating/cooling options, while a promising decrease in thermal discomfort in all zones in both seasons. | [75] |
| 2014 | NN-based approach with a MAS infrastructure to improve energy efficiency, while maintaining acceptable thermal comfort level for occupants of a UCLan's building | Recorded temperature and sensor metering data collected from an actual building/Interviews | Combining a gaussian adaptive resonance theory map (gARTMAP) with MAS for building-IHMS | Distributed AI | Learning the user's thermal preferences | Current date and time; Outdoor temperature; Room temperature; Temperature of the heating element of the radiator; Hot/Water temperature; Desired room temperature (Human input) | Simulation results showed that the proposed gARTMAP-MAS IHMS might use less heat to achieve the desired indoor temperature, compared to the existing rule-base BMS and fuzzy ARTMAP IHMS | [76] |
| 2014 | Control logic for thermally comfortable and energy-efficient environments in buildings with double skin envelopes | Weather data/Building model data/Indoor thermal conditions data | Rule-base control logic and ANN-based control logic development for openings and cooling systems in summer | Predictive and adaptive control | Comfort range built from the cavity and indoor temperature conditions | Cavity air temperature; Indoor/Outdoor air temperature; Opening conditions of the envelope | ANN-based logic showed significant results in reducing over/undershoots out of the comfort range. Also, using simplest rule-base control logic allowed to save cooling energy. | [77] |

**Table 1.** Cont.

| Year | Study Case | Source of Data used | Underlying AI/ML techniques | Application Scenario | Thermal comfort-based model/method | Input(s) & Controlled parameter(s) | Key Results | Ref |
|---|---|---|---|---|---|---|---|---|
| 2014 | Stochastic optimized controller development to improve the energy consumption and indoor environmental comfort in smart buildings | Simulations data: occupants' data/Outdoor information | Multi-agent control system combined with GA development to find the optimal set-points | Distributed AI and optimized setting | Temperature set-point defined by user's preferences (via user interface) | Customer preferences; Temperature; Illumination level; $CO_2$ concentration | By defining comfort ranges as constraints, the overall occupant comfort with GA has embedded was kept between 0.97 and 0.99, and the error between set-points and the sensor data became smaller with GA. A significant reduction in the overall energy consumption (~20% compared to system without GA) | [78] |
| 2015 | Developing and testing an NN-based smart controller for maintaining a comfortable environment, and thus saving energy using a single zone test chamber | Indoor climatic data collected from a test chamber conducted in a university campus located in Glasgow | Random NN-based controller development and trained using the hybrid PSO with sequential quadratic programming | Predictive control and optimized setting | User recommendations /PMV-based set-points (Fanger's model) | Occupant preferences; Room air temperature; Air inlet temperature; $CO_2$ concentrations in HVAC dust/room; Actuation signal of inlet air | The proposed controller has learned the human preferences with an accuracy of 94.87% for heating, 98.39% for cooling and 99.27% for ventilation. The occupancy estimation using RNN is about 83.08%. | [79] |
| 2015 | Predictive-based controller development for multizone HVAC systems management in non-residential buildings | Climatic and occupancy data measurement of a non-residential building located in Perpignan in France | Low-order ANN-based models' development (as controller's internal models) to supervise the HVAC subsystems and tuned through GA to solve the optimization problems | Predictive control and optimized setting | PMV (Fanger's model) | Air temperature; Radiant temperature; Room occupancy; | The proposed strategy allowed to optimize the operation times of HVAC subsystems by computing the right time of turning on/off, while reducing energy consumption and improving significantly thermal comfort for cooling/heating modes and year period, compared to the basic scheduling approaches. | [80] |
| 2015 | AI-theory-based optimal control for improving the indoor temperature conditions and heating energy efficiency of the building with double-skin | Weather data from TMY2 of Seoul, South Korea/Computer simulation datasets using MATLAB and TRNSYS/Building test-model data | AI-theory-based optimal control algorithms development including ANN, FL, ANFIS with 2 input, and ANFIS with one input, for improving the indoor temperature conditions and heating efficiency | Five control algorithms including: Rule + ANN; ANN + ANN; Fuzzy + ANN; ANFIS with 2 inputs + ANN; ANFIS with 1 input + ANN | Defined comfort temperature range | Indoor air temperature; Cavity air temperature (of the double skin); Outdoor air temperature; Surface opening status | Compared to the rule-based algorithm, FL, ANFIS-2 inputs and ANFIS-1 input models increased significantly the comfortable condition period by 2.92%, 2.61% and 2.73% resp. When heating energy efficiency was the main interest, then the ANN-based algorithm is applicable by reducing the SD from the average and 0.5 to 80.34 and 56.00% resp. | [81] |
| 2015 | Hybrid predictive control model development for energy and cost savings in a commercial building (Adelaide airport) | Data collected from a building management system-Johnson Controls Australia Pty Ltd/ Meteorological data obtained from the Bureau of Meteorology of Australia | Combining a linear MPC with a neural network feedback linearization (NNFL) for energy and cost savings | Model-based predictive control | Comfort range defined by ASHRAE 55: Indoor temperature ∈ [21.5°C, 24°C] during occupancy hours | Supply air temperature; Chilled water temperature; | By maintaining the indoor temperature within the defined comfort range [21.5°C – 24°C] during occupancy period (from 5:00 am to 9:30 pm), simulation results showed that about 13% of energy cost saving was achieved and up to 41% of energy saving, compared to the baseline control. | [82] |
| 2015 | Agent-based particle swarm optimization development for inter-operation of Smart Grid-BEMS framework | Data measurements from: a feeder of a Dutch low voltage network with 74 customers/connection point of a 3-floor office building, Weather data of a winter day in the Netherlands, Occupancy profiles | Agent-based control scheme + PSO for maximizing comfort and energy efficiency | Distributed AI and optimized setting | Comfort was modeled as a temperature Gaussian function | Occupancy information; Indoor temperature; Indoor relative humidity; $CO_2$ concentration; Measured weather data | The proposed system could effectively improve the voltage profile of the feeder, while ensuring acceptable comfort levels. | [83] |

| Year | Study Case | Source of Data used | Underlying AI/ML techniques | Application Scenario | Thermal comfort-based model/ method | Input(s) & Controlled parameter(s) | Key Results | Ref |
|---|---|---|---|---|---|---|---|---|
| 2015 | Fuzzy logic-based advanced on–off control for maintaining thermal comfort in residential buildings | Temperature data measurements (using sensors)/ Weather data of the Republic of Korea | On–off controller combined with fuzzy algorithm for thermal comfort | Fuzzy logic-based control | Desired room temperature was set to 20ºC | Room air temperature **Building parameters:** Indoor air heat capacity; Floor heat capacity; Ceiling heat capacity; Wall heat capacity; Window heat capacity; Door heat capacity; Equivalent diameter; Boltzmann constant; Wall height; Floor dimension; Ceiling dimension; Wall/Window/Door dimensions; | Compared to the conventional on–off controller, the proposed system had better control performance and saved energy. | [84] |
| 2015 | Automatic air-conditioning control development for indoor thermal comfort based on PMV and energy saving | Experiment data: Indoor climatic (environmental sensors) and personal data/Questionnaires | ANFIS + particle swarm algorithm (PSA) used to solve the inverse PMV model and determining comfort temperatures | 4 control methods: Fixed temperature setting; Inverse PMV mode with FF-PID; Inverse PMV mode with FF-Fuzzy; Inverse PMV mode with self-tuning control | Inverse-PMV model based on the desired PMV and measured air-velocity and humidity | Indoor air temperature; Relative humidity; Air velocity | The proposed control method performed better than conventional method by effectively maintaining the PMV within a range ±0.5 and up to 30% of energy saving. | [85] |
| 2015 | Implementing and evaluating a multi-grid reinforcement learning method for energy conservation and comfort control of HVAC systems in buildings | Office building profile SmOffPSZ provided by EnergyPlus/Weather data of Summer daytime period of Beijing by EnergyPlus | A multi-grid method for Q-leaning development for HVAC control optimization | Optimized setting | PPD-PMV (Fanger's model) | Outdoor temperature; Indoor temperature; Indoor humidity | Simulation results showed that the proposed multi-grid approach helped to accelerate the convergence of Q-learning, and performed better on energy saving and comfort than the constant grid versions. | [86] |
| 2015 | Multi-agent control architecture for cooling and heating processes in smart residential building. | Weather data from SEA; Environmental data from physical sensors; Occupancy data at home and at work; Human behavior data: interviewing 5 volunteers to record their activities using smart phone logger at home & RFID system at work) | Multi-agent control system (MACS) combined with ML algorithms for occupancy prediction | Machine learning and distributed AI | Desired temperature based on occupant's behavior | Outdoor temperature; Indoor temperature; Occupancy at home; Occupancy at work; Heater power rate | The proposed system allowed to significantly improve the occupants comfort with a slight increase in energy consumption, with respect to 'sense behavior', compared to simple strategies with predefined temperatures | [87] |
| 2016 | Simulation-based MPC procedure for multi-objective optimization of HVAC system performance and thermal comfort, applied to a multi-zone residential building in Naples, Italy. | Climatic conditions data taken from IWEC data file for Naples/ Occupancy profiles provided by IWEC [88] | MPC + GA optimization for the best solutions for HVAC system control in a day-ahead horizon | Prediction control and optimized setting | $PPD^{MAX}$: the maximum hourly value of PPD (Fanger's model) | Weather conditions; Occupancy profiles | Up to 56% operating cost reduction and improvement in thermal comfort, compared to the standard control strategy. | [89] |
| 2016 | Simulation-based multi-objective optimization for building energy efficiency and indoor thermal comfort | Wall and glazing specifications data were based on EnergyPlus (ASHRAE materials) databases/Average solar absorptance data obtained from [90]/ Weather data from the national center of the climatology of Iran | Implementing a multi-objective artificial bee colony (MOABC) optimizer to minimize the total energy consumption and the predicted percentage of dissatisfied | Optimized setting | PPD (Fanger's model) | **Continuous variables:** Room rotation; Window size; Cooling/heating setpoint temperatures; Glazing material features; Wall conductivity **Discrete variables:** Thermal, solar and visible absorptance coefficients of the wall; Solar and visible transmittance coefficients of the glazing | The multi-objective optimization coupled with TOPSIS decision-making method showed that in different climates, even though the energy consumption increased a bit by 2.9-11.3%, the PPD significantly reduced by 49.1-56.8%, compared to the baseline model. | [91] |

| Year | Study Case | Source of Data used | Underlying AI/ML techniques | Application Scenario | Thermal comfort-based model/ method | Input(s) & Controlled parameter(s) | Key Results | Ref |
|---|---|---|---|---|---|---|---|---|
| 2016 | An operation collaborative optimization framework development for a building cluster with multiple buildings and distributed energy systems while maintaining indoor thermal comfort | Three DOE reference small office (post-1980)/medium office (post-1980)/real small size commercial buildings were used [92] | Multi-objective optimization through PSO was used to determine the framework operation strategies | Optimized setting | PMV (Fanger's model) | Temperature; Battery charging/discharging currents; Ice tank | Around 12.1–58.3% of energy cost saving under different electricity pricing plans and thermal comfort requirements. | [93] |
| 2016 | ANN-based algorithms development for optimal application of the setback moment during the heating season | Datasets were collected in the test module located in Seoul, South Korea /Meteorological TMY2 data were used for test location | ANN model was developed to predict the optimal start moment of the setback temperature during the normal occupied period in a building | Predictive control and optimized setting | Defined set-point temperature (20–23°C based on indoor and outdoor temperatures) for occupied periods | Indoor temperature; Outdoor temperature; Temperature difference from the setback temperature | The optimized ANN model showed a promising prediction accuracy by a R-squared up to 99.99%. The developed ANN-based algorithms were much better in thermal comfort improvement (97.73% by Algorithm (1)) or energy saving (14.04% by Algorithm (2)), compared to the conventional algorithm. | [94] |
| 2016 | ANN-based control algorithm development for improving thermal comfort and building energy efficiency of accommodation buildings during the cooling season | Climatic conditions data collected during cooling season/ Datasets collected from the simulation model (using MATLAB and TRNSYS) | Two ANN-based algorithms: 1st model for predicting the cooling energy consumption during the setback period and 2nd model for predicting the optimal starting moment of thermal control during setback periods | Predictive and adaptive controls | Fixed set-point temperature during occupied periods (23°C with 3°C dead-band) and setback temperature (25°C) for unoccupied hours | Indoor air temperatures; Outdoor air temperature | Simulation results showed that ANN models gave accurate prediction results with acceptable error (for thermal comfort and energy saving improvement): 1st model: Average difference = 17.07%/MBE = 17.66%, 2nd model: Average difference = 20.87%/MBE = 21.90% | [95] |
| 2016 | Multi-objective control and management for smart energy buildings | Interviews/Indoor and Outdoor climatic data collected using physical sensors (during the experiments) | Hybrid multi-objective genetic algorithm (HMOGA) development for optimizing the energy management | Optimized setting | Discomfort parameter based on the user preferences, defined as: $Discomfort_{Temp} = \left(\frac{Temp - Temp_{set}}{Temp_{set}}\right)^2$ | Temperature; Illumination level; $CO_2$ concentration | 31.6% energy saving could be achieved for smart control building, and the comfort index was improved by 71.8%, compared to the conventional optimization methods. | [96] |
| 2016 | Real-time information-based energy management controller development for smart homes applications | Data collection using physical sensors (human occupancy)/ Simulation data | GA was used for solving the complex energy optimization and appliance scheduling problem | Optimized setting | Customer preferences | User preferences; External parameters (price signal, user presence, temperature) | The proposed algorithms are flexible enough to maintain the user's comfort while reducing the peak to average ratio (PAR) and electricity cost up to 22.77% and 22.63% resp. | [97] |
| 2016 | Deploying an intelligent MBPC solution for HVAC systems in a University building | Data collection (6768 samples): Atmospheric data collected by an intelligent weather station [98]/Room data collected by SPWS devices/HVAC data using BMS interface software | A MOGA framework was used to design the predictive model radial basis function (RBF) neural networks (NN) | Predictive control | PMV (Fanger's model) | Room air relative humidity; Room air temperature; Air conditioning reference temperature; Atmospheric air relative humidity; Solar radiation | The IBMPC HVAC showed significant results in reducing energy cost and maintaining thermal comfort level during the whole occupation period, compared to scheduled standard, temperature-regulation control. | [99] |
| 2016 | A personalized energy management system (PEMS) development for HVAC systems in residential buildings | Experiments Data collected from a laboratory building during 3 months (8:30 AM to 10:00 PM) (using sensors)/ Weather forecast were obtained from the internet | Hidden Markov model (HMM) used for modelling occupancy + ANFIS for modeling the occupant behavior | Predictive control | comfort margins specified by the user using a thermostat (i.e., personalized comfort bands) | Ambient temperature; Room temperature; Occupancy | By maintaining temperature within the comfort band, about 9.7% to 25% reduction in energy consumption as well as the cost, from 8.2% to 18.2%. | [100] |

| Year | Study Case | Source of Data used | Underlying AI/ML techniques | Application Scenario | Thermal comfort-based model/ method | Input(s) & Controlled parameter(s) | Key Results | Ref |
|---|---|---|---|---|---|---|---|---|
| 2017 | Proposing an AI-based heating and cooling energy supply model, responding to abnormal/abrupt indoor situations, to enhance thermal comfort and energy consumption reduction | Climatic, geometries and human data for each building were adopted from references [101], [102] | A decision making based-ANN model was developed to maintain desired room temperature and optimize cooling and heating air supply | Optimized setting | PMV-PPD (Fanger's model) | Relative humidity; Heat loss; Room temperature **Human factors/indoor conditions:** Metabolic rate; Clothing insulation; Cooling/Heating temperatures | Thermal comfort improvement by 2.5% for an office building, and around 10.2% for residential buildings, as annual energy consumption reduction by 17.4% for an office building and 25.7% for residential buildings. | [103] |
| 2017 | AI-based controller development for improving thermal comfort and reducing peak energy demands in a district heating system | Temperature information of the past 62 years in Seoul, South Korea/Other climatic, geometries and users' data were adopted from references [101], [102] | ANN-based thermal comfort optimizer (Opt. ANN) development | Optimized setting and predictive control | PMV-PPD (Fanger's model) | Relative humidity; External work rate; Air velocity; Clothing insulation; Metabolic rate; Temperature; Floor area; Building height; Wall/ Window depth | The proposed model's effectiveness was up to 27% for thermal comfort, and a reduction of annual energy loss over 30% for cooling and 40% for heating, compared to a conventional thermostat on/off controller. | [104] |
| 2017 | A personalized thermal comfort model (BCM) development for smart HVAC systems control | The ASHRAE RP-884 dataset [105]/ Experimental data collected by the authors at the University of Southampton in UK (interviews, sensors measurements)/Thermal properties of chosen houses and HVAC systems are based on data presented by [106]. | Bayesian network-based model development for learning individual users' preferences | Predictive control and optimized setting | BCM by combining the static and the adaptive models to quantify: the user's optimal comfort temperature ($T_{opt}$); the user's vote ($T_{vote}$); and the user's thermal sensitivity ($\gamma_v$) | Indoor climate conditions; User votes; Outdoor weather conditions; Number of occupants during summer/winter; Observation count for each occupant; | By using an alternative comfort scale, the proposed model outperformed the existing approaches by 13.2%–25.8%. The heating algorithm allowed to reduce energy consumption by 6.4% to 10.5% for heating, and by 15.1% to 39.4% for air-conditioning, while reducing discomfort by 24.8%. | [107] |
| 2017 | A low-cost, high-quality decision-making mechanism (DMM) targeting smart thermostats in a smart building environment located in Chania, Greece | Building specifications based on real buildings located in Greece/ Weather conditions data and Energy pricing data for 2010 [108]. | ANN + Fuzzy Inference System (FIS) combination to determine temperature set-points/thermal zone and their dynamic refinement | ANN + Fuzzy logic (FIS) | PPD (Fanger's model) | Thermal zones' temperature; Humidity; Number of occupants per room; Current external weather conditions (Temperature; Humidity; Solar radiation) | Comparing to RBCs, the proposed framework allowed to reach a higher thermal comfort while reducing energy consumption by an average between 18%–40%. The use of FL and considering the dynamic behavior of the world allowed to improve the total cost by 7%–19% on average. | [109] |
| 2017 | Designing and implementing a smart controller by integrating the internet of things (IoT) with cloud computing for HVAC within an environment chamber. | Occupancy and Indoor climatic data were measured using sensor nodes (by the authors). | Random neural network (RNN) model development for occupancy estimation, and optimized with the hybrid PSO-SQP | Occupancy estimation-based control and optimized setting | PMV (Fanger's model) | HVAC inlet air temperature; HVAC inlet air $CO_2$ concentrations; Inlet air temperature of the environment chamber; $CO_2$ concentration of the environment chamber | By maintaining the PMV set-points, results showed that the hybrid RNN-based occupancy estimation algorithm was accurate by 88%. About 27.12% reduction in energy consumption with the smart controller, compared to the simple rule-based controllers. | [110] |
| 2017 | RNN-based smart controller development for HVAC by integrating IoT with cloud computing and web services | Occupancy and Indoor climatic data were measured using sensor nodes (by the authors) | RNN models were trained with PSO-SQP for estimating the occupancy and PMV set-points for HVAC control | Optimized setting and occupancy estimation | PMV (Fanger's model) | Room temperature; Inlet air temperature; Inlet $CO_2$ concentration; Indoor $CO_2$ levels; Inlet air actuation signal (valve opening) | By evaluating the intelligent controller architectures, the energy consumption was 4.4% less than Case-1 and 19.23% less than Case-2. The RNN HVAC controller was successfully able to maintain the user defined set-points and accurate temperature for PMV set-points. | [111] |

**Table 1.** Cont.

| Year | Study Case | Source of Data used | Underlying AI/ML techniques | Application Scenario | Thermal comfort-based model/ method | Input(s) & Controlled parameter(s) | Key Results | Ref |
|---|---|---|---|---|---|---|---|---|
| 2017 | A newly developed Epistemic-Deontic-Axiologic (EDA) agent-based solution supporting the energy management system (EMS) in office buildings | Experimental studies were conducted in an (air-conditioned) academic building by applying standardized questionnaire surveys [112] to collect subjects' personal data/Climatic data recorded by sensors | Support vector machine (SVM) based algorithms: SVM and C-SVC was embedded to BEMS to establish thermal sensation model and comfort requirement | Distributed AI and machine learning | Personal thermal sensation model for MET is occupant's personal activity: $TSV = F_p(T_a, \bar{T}_r, V_a, RH, MET, Clo)$ Group-of-people-based thermal sensation model for MET is group of people average activity: $AMV = F_g(T_a, \bar{T}_r, V_a, RH, MET, Clo)$ | Temperature; Relative humidity; Globe temperature; Air velocity; Occupant's personal activity; Clothing insulation level | Case studies simulations showed the abilities of the developed model in energy saving by 3.5–10%, compared to the pre-set control systems, while fulfilling the individual thermal comfort requirements (by maintaining the average value of TSV within the range [-0.5, +0.5]) | [113] |
| 2017 | Deploying a software application based mobile sensing technology (Occupant Mobile Gateway (OMG)) for occupant-aware energy management of mix of buildings in California | Pilot test-sites conducted in 4 different mix-academic buildings in California: Subjective feedback data using OMG App. And objective thermal data measurements using embedded sensors/Data vintages (pre-1980, post-1980 and ASHRAE 90.1) from the US-DOE medium size office building as reference model [114] | Logistic regression (LR) techniques were applied for generating personalized comfort profiles and group-level models | Machine learning and predictive control | Generating data-driven thermal comfort model by learning from real-time occupant subjective feedback via smart-phone/server (OMG) application and objective thermal data | Indoor temperature; Relative humidity; Location; Occupancy; Subjective feedback | Simulations results implementing occupant-driven models showed that thermal management learned by subjective feedback had the potential energy savings while maintaining acceptable levels of thermal comfort | [115] |
| 2017 | Implementing a predictive control strategy in a commercial BEMS for boilers in buildings | Experimental data collected from two heating seasons through a BEMS having a set of 22 temperature sensors implemented in an academic building located in Spain. | NN development to predict time required for building conditioning | Predictive control | Predefined internal temperature: an average of 20°C at 8:00 and 22°C throughout the rest of the day | Internal temperature; External temperature; Water heating system temperature | The predictive strategy allowed to reduce around 20% of energy required to heat the building without compromising the user's thermal comfort, compared to BEMS based on scheduled ON/OFF control. | [116] |
| 2017 | An HVAC optimization framework deployment for energy-efficient predictive control for HVAC systems in office buildings | Real data measured using sensors and meters: Climatic and occupancy levels data in an academic building located in Spain | Random forest (RF) regression techniques used for energy- efficient predictive control of HVAC | Predictive control and optimized setting | Comfort ranges defined by Royal Decree 1826/2009, i.e., setting indoor temperature between 21°C and 26°C | Indoor/Outdoor temperatures; Indoor/Outdoor relative humidity; Occupancy level | The proposed Next 24h-Energy framework showed significant results in reducing energy consumption for heating (48%) and cooling (39%), without affecting the user's comfort (defined by indoor temperature between 21°C and 26°C). | [117] |
| 2017 | A smart heating set-point scheduler development for an office building control located in the UK | Building information model located in Cardiff, UK/ Occupant surveys (N=30)/ Weather information from local weather stations/Data generated using EnergyPlus by varied heating set-points. | A multi-objective GA was coupled with ANN model for energy sum and average PPD (for occupied hours) calculation during 24-hour period | Predictive control and optimized setting | PPD (Fanger's model) | Weather information (Outdoor temperature; Solar radiation; Humidity); Hour of the day; Heating set-point temperature; Occupancy profile; Indoor temperature (of the previous 3 times steps) | 4.93% energy savings whilst improving thermal comfort by reducing the PPD by an average of 0.76%. | [118] |

Table 1. Cont.

| Year | Study Case | Source of Data used | Underlying AI/ML techniques | Application Scenario | Thermal comfort-based model/ method | Input(s) & Controlled parameter(s) | Key Results | Ref |
|---|---|---|---|---|---|---|---|---|
| 2017 | A deep reinforcement learning based data–driven approach development for building HVAC control | Weather profiles obtained from the National Solar Radiation Data-Base [119] and time-of-use price data from the Southern California Edison [120]/ EnergyPlus models data. | Deep reinforcement learning (DRL)–based algorithm | Optimized setting | Desired temperature range [19°C, 24°C] based on ASHRAE standard | Current (physical) time; Zone temperature; Environment disturbances (Ambient temperature; Solar irradiance intensity; Multi-step forecast of weather data) | Up to 20%-70% energy cost reduction while meeting the room temperature requirements, compared to a conventional rule-based approach. | [121] |
| 2017 | A reinforcement learning-based thermostat schedule controller development using long–short–term memory recurrent neural network for an office HVAC system | Simulation data from a single office space model (EnergyPlus/MATLAB/BCVTB)/Occupancy data measurements (3 occupants) Weather data | Actor–critic–based RL and Long-Short-Term Memory (LSTM) recurrent neural network (RNN) | Optimized setting | PMV (Fanger's model) | Office occupancy; Indoor/Outdoor temperature; Solar irradiance; Cloud cover; Energy demand from the last time step | An average 2.5% energy savings was achieved while improving thermal comfort by an average of 15%, compared to other control baselines (Ideal PMV & Control Variable). | [122] |
| 2017 | A hybrid rule-based energy saving approach development using ANN and GA in buildings. | Data obtained from authors' simulation model/Historical, warnings and recommendations data recorded via a GUI/Climatic and occupancy data collected using installed sensors in the pilot zone | ANN model used with GA-based optimization process for generating optimal energy saving rule | Optimized setting | PMV (Fanger's model) | Time information; Outdoor air temperature; Wind speed; Wind direction; Solar radiation; Solar azimuth angle; Solar altitude angle; Zone air temperature; Zone air sensible heating rate; Zone ideal total cooling rate; Occupancy | Validation results showed an average 25% energy savings while satisfying occupants' (elderly people) comfort conditions, i.e., PMV was kept within the range of [-1, +1]. | [123] |
| 2017 | Deploying machine learning techniques to balance energy consumption and thermal comfort in ACMV systems through computational intelligence techniques in optimizations | Experimental (environmental and personal) data collected in the thermal laboratory of an academic building (from a previous study conducted by the authors) [124] | Extreme learning machines (ELM) and NN models were integrated with sparse Firefly Algorithm (sFA) and sparse Augmented Firefly Algorithm (sAFA) | Predictive control and optimized setting | PMV (Fanger's model) | **Environmental parameters:** Ambient air temperature; Ambient air velocity; Air relative humidity; Mean radiant temperature / **Occupant parameters:** Metabolic rate; Clothing insulation factors; External work done | Maximum energy saving rate (ESR) prediction was about -31% using sparse AFA optimizations while maintaining thermal comfort within the pre-established comfort zone (when PMV~0). | [125] |
| 2017 | Machine learning-based thermal environment control development | Climatic and Subjective data from human thermal comfort experiments conducted in environment chamber in an office building in California | ANN-based algorithm used for predicting the occupant's thermal preference | Prediction control | Individual's thermal preference/feedback | Environment temperature; Relative humidity; Thermal comfort feedback | A total of up to 45% more energy savings and 44.3% better thermal comfort performance than the PMV model. | [126] |
| 2018 | The benefits of including ambient intelligent systems for building's EMS control to optimize the energy/comfort trade-off | Occupants' subjective votes by varying temperature by 1°C/Occupancy data collection using RFIDS cards/Other collected data: Current indoor temperature and time of the day measured by the authors | k-means algorithm enabling automatic configuration of HVAC system | Optimized setting | Learning occupants' preferences (via individual subjective rating votes) to quantify the group occupant comfort | User's vote; User's presence | The energy consumption was reduced by an average of 5KWh while maintaining the majority of the occupants within acceptable comfort levels (the comfort rate was 5% lower than the baseline). | [127] |
| 2018 | Optimizing the passive design of newly-built residential buildings in hot summer and cold winter region of China | Weather data was a TMY of Shanghai from the EnergyPlus website [128]/Dataset with a sample of 1100 cases generated by SimLab software/Construction data of the base-case building model based on a real apartment in Shanghai, some features were adopted from the building code – DGJ [129] | Non-dominated sorting GA II (NSGA-II) was combined with ANN model for multi-objective optimization | Optimized setting | The annual indoor thermal CTR [%] and DTR [%] based on Szokolay's theory: $CTR = \frac{1}{n} \times \sum_{i=1}^{n} \frac{N_i}{8760} \times 100$  $DTR = 100 - CTR$ | 37 variables related to natural ventilation, solar shading, thermal insulation and passive solar heating | The defined objectives Comfort Time Ratio (CTR) and Building Energy Demand (BED) were significantly improved, i.e., the annual thermal comfort hours were extended by 516.8–560.6 hours, and the annual BED was reduced by 27.86–33.29% compared to base-case design. | [130] |

Table 1. Cont.

| Year | Study Case | Source of Data used | Underlying AI/ML techniques | Application Scenario | Thermal comfort-based model/ method | Input(s) & Controlled parameter(s) | Key Results | Ref |
|---|---|---|---|---|---|---|---|---|
| 2018 | A novel type of decentralized and cooperative method development for decision-making strategies in the buildings' context, based on reinforcement learning. | Weather data of 2013 were from the KNMI, Construction data from DOE 2004 standard for the Netherlands were used for all considered buildings, Climatic conditions and Occupancy levels measurement for each building | Extended joint action learning (eJAL) was developed and compared with Q-learning and Nash n-player game (Nash-NPG) methods | Optimized setting | Thermal comfort index is conceptualized as a Gaussian function of indoor temperature $\xi e\left[\frac{-(T-\mu_T)^2}{2\sigma_T^2}\right]$ | Outdoor conditions (Outdoor temperature; Relative humidity); Occupancy levels; Lights; Device usage; Thermostat profiles | The long-term learning analysis showed that Q-learning and eJAL gave acceptable comfort losses ($\Delta C \leq 0.4$), for demand/response balance, eJAL (Median=1.67) was slightly better than Q-learning (Median=2.21) | [131] |
| 2018 | Plug&play solution of an HVAC thermostat's set-point scheduling inspired by reinforcement learning technique | Building dynamic and sensor data were produced by the EnergyPlus suite [108]/Building model data from an actual building located in Greece/Weather data collected in 2010 were publicly available | Neural Fitted Q-iteration (NFQ)-RL based algorithm deployment for control performance | RL-based controller development | PMV (Fanger's model) | Outdoor temperature; Solar radiation; Indoor humidity; Indoor temperature | With energy/comfort trade-off balance, an average up to 32.4% energy savings and up to 27.4% comfort improvements in average, compared with rule-based control set-points. | [132] |
| 2018 | A demand-driven cooling control (DCC) based on machine learning techniques for HVAC systems in office buildings | Building construction data and thermal environment comply with the Green Mark Platinum standard [133]/ Occupancy levels data collected using motions sensors installed in 11 offices for 7.5 months, Other measured data: Room climate (via sensors)/ Interviews through HMI/Energy usage data recorded (energy meters) | k-means clustering and kNN algorithms were applied for learning the occupants' behavior | Machine learning and predictive control | Predefined comfort conditions (comfort mode set-points: 22°C and 22.5°C) during working hours (8:00 to 18:00)/weekdays | Occupancy profiles (time of the daily first arrival/last departure; daily maximum vacancy duration during working hours); Indoor air temperature; Indoor relative humidity; Indoor $CO_2$ concentration; Occupants' preferences | While maintaining room temperature to the comfort set-point (temperature deviations means all less than 0.1°C in both the baseline and the DCC tests); between 7% and 52% energy savings were ensured compared to the conventionally-scheduled cooling systems. | [134] |
| 2018 | A novel real-time automated HVAC control system built on top of an Internet of Things (IoT) | Experiment data: Climate conditions using sensors/ Interviews (User-feedback interface)/Occupancy tracked by embedded sensors | ANN MPL-based times-series predictive model + Mixed Integer Linear Programming (MILP) problem optimizer | Predictive control and optimized setting | User's zoning feedback reflecting his dissatisfaction, while thermal comfort is function of temperature based on ISO 7730: $\phi_n = 1 - \left(\frac{|t_n^{room}-t_n^{set}|}{dev_n}\right)$ | User's feedback; Indoor thermal parameters (temperature; relative humidity; brightness; $CO_2$ level; air pressure; smelling) | Between 20% and 40% energy savings were achieved while maintaining temperature within the comfort range (except the pre-peak cooling hour). | [135] |
| 2018 | Agent-based control system for and optimized and intelligent control of the built environment | Simulation data | An evolutionary MOGA development for achieving energy-comfort trade-off | Distributed AI and optimized setting | Thermal comfort ranges based on the users' preferences as $[C_{min}, C_{max}]$ within the ASHRAE standard | Users' preferences; Temperature; Artificial illumination; $CO_2$ concentrations in air; Relative humidity | By applying MOGA optimizer allowed to save up to 67% energy consumption and about 99.73% overall comfort improvement. | [136] |
| 2018 | Combining a Comfort Eye sensor with a sub-zonal heating system control for building climate management | Test room (office building in Italy) characteristics data/ Climate data measurement using the 'Comfort Eye' system/ Outdoor temperature collected from a local weather station. | PID tuned with fuzzy logic (PID-PMV) | Fuzzy PID | PMV (Fanger's model) | Mean radiant temperature; Wall temperature; Air temperature; Relative humidity; Air velocity; PMV | Up to 17% energy savings with respect to the standard ON/OFF mono-zone control, thermal comfort has been slightly improved with a minimum deviation from the neutral condition (PMV=0) | [137] |
| 2018 | A whole BEM-DRL framework development for HVAC optimal control in a real office building (Intelligent Workplace (IW)) located in Pennsylvania | Building specifications data/Three months measured climate data (from Jan 1st 2017 to March 31th 2017) / TMY3 weather data [108] | A DRL-based model development | Optimized setting | PPD (Fanger's model) | Mullion system supply water temperature set-point **Weather conditions:** Outdoor air temperature; Humidity; Solar radiation; Wind speed; Wind direction | About 15% heating energy savings with similar comfort conditions as the base-case | [138] |

| Year | Study Case | Source of Data used | Underlying AI/ML techniques | Application Scenario | Thermal comfort-based model/ method | Input(s) & Controlled parameter(s) | Key Results | Ref |
|---|---|---|---|---|---|---|---|---|
| 2019 | An indoor-climate framework development for air-conditioning and mechanical ventilation (ACMV) systems control in buildings | Environmental parameters were recorded during the experiments using embedded sensors/ Subjective data were obtained through interviews/ A sample of 1155 energy models' data was experimentally collected | Single layered feedforward artificial neural network (SLFF-ANN)-based energy model + OAT optimization algorithm | Predictive control and optimized setting | TSI (*Cool-Discomfort/Comfort/Warm-Discomfort*) based on ASHRAE 7-point sensation scale | **ACMV system:** Energy; Frequency **Environmental:** Air temperature; Relative humidity; Mean radiant temperature/ **Physiological:** Skin temperature; Pulse rate; Skin conductance; Oxygen saturation/ **Personal:** Clothing insulation; Body surface area; Age | An average of 36.5% energy saving was ensured, and it was found that 25°C is the ideal comfort temperature with a minimum energy use. | [139] |
| 2019 | A novel optimization framework using a deep learning-based control for building thermal load | Data from a reference office building at PNNL/Weather information from the TMY-3/ Utility rate structure from schedule E-20 tariff of the Pacific Gas and Electric Company [140] | Recurrent neural network (RNN) development for thermal load prediction + Black-box optimization (Mesh Adaptive Direct Search (MADS)) | Load prediction and optimized setting | Defined zone temperature cooling set-points (23.3°C in occupied hours) | Weather information; Occupancy states | Up to 12.8% cost savings compared with a rule-based strategy, while maintaining the users' thermal comfort during the occupied periods. | [141] |
| 2019 | A learning-based optimization framework development for HVAC systems in smart buildings | 15,000 hours of simulation data in TRNSYS were used for training and testing performances/Weather dataset from SG-Singapore-Airp-486980 [142] | Deep NN for predicting thermal comfort + deep deterministic policy gradients (DDPG) for energy optimization | Bayesian predictive control and optimized setting | Predicted thermal comfort value at time slot $t$: $M_t = \Phi(T_t^{in}, H_t^{in})$ | **Thermal control model parameters:** Indoor/Outdoor air temperature; Indoor/Outdoor air humidity **Comfort prediction parameters:** Air temperature; Humidity; Radiant temperature; Air speed; Metabolic rate; Clothing insulation | DDPG allowed to achieve higher degree of thermal comfort with an average value closer to the preset threshold of 0.5. DDPG could save 6% more energy than the baseline methods. | [143] |
| 2019 | AI-based agent development for indoor environment control while optimizing energy use of air-conditioning and ventilation fans in a classroom and a laboratory | Experimental data (Climate, energy, subjective and occupancy) collected using embedded sensors in a laboratory room and a classroom environment/ Weather information (10th years data) from EnergyPlus [144] | A deep-RL (with double Q-learning) algorithm was adopted | Optimized setting | PMV (Fanger's model) | Indoor temperature; Outdoor temperature; $CO_2$ levels; PMV index | AI-agent has successfully managed the indoor environment within an acceptable PMV values between -0.1 and +0.07, and 10% lower $CO_2$ levels, while reducing energy consumption by about 4% to 5%, compared with a conventional control system | [145] |
| 2020 | A novel MPC relied on artificial intelligence-based approach development for institutional and commercial buildings control. | Hourly measurements from October 1st, 2017 to March 31st, 2018 of an institutional building; Weather forecasts retrieved using CanMETEO [146] | Gauss Process Regressions (GPR) model with squared exponential Kernel function applied to MPC for control-oriented model development | Predictive control and optimized setting | Pre-defined set-point ramps (temperature) profiles | Outdoor air temperature (OAT); Indoor air temperature set-point ($T_{sp}$); Indoor air temperature set-point variation ($\Delta T_{sp}$); Predicted building heating load (HD); Electrical baseload (EBL); Electric demand margin (EDM); Total heating load (HL); Daytime value; Occupancy level | The AI-based MPC strategies allowed to reduce the natural gas consumption and the building heating demand by 22.2% and 4.3% resp. as well as improving thermal comfort, while minimizing the required amount of time and information, compared with *business as usual* control strategies. | [147] |

**Table 1.** Cont.

| Year | Study Case | Source of Data used | Underlying AI/ML techniques | Application Scenario | Thermal comfort-based model/ method | Input(s) & Controlled parameter(s) | Key Results | Ref |
|---|---|---|---|---|---|---|---|---|
| 2020 | Hybrid data-driven approaches development for predicting building indoor temperature response in VAV systems. | Data generated using EnergyPlus building simulation models: 1) One-story office building located in Newark, NJ and 2) DOE medium office building located in Chicago, IL. | Multivariate linear regression (MLR) and ANN trained by Bayesian Regulation (BR) algorithm models were used for predicting the indoor temperature variation | Predictive control | Defined comfort zones | Damper position variables; ASHRAE sky clearness index; Lighting schedule; Equipment schedule; Occupancy schedule; Zone average temperature; Wall surface temperature; Dry/Wet-bulb external temperature; Supply air temperature; Outside air velocity; Time-of-day, Incident solar radiation | The proposed model allowed to improve the control and optimization of buildings space cooling | [148] |
| 2020 | An event-triggered paradigm based on RL approach for smart learning and autonomous micro-climate control in buildings. | Generated data using a real Building model data in EnergyPlus/EnergyPlus Chicago Weather data (Chicago-OHare Intl AP 725300) is used for simulation. | Stochastic and deterministic policy gradient RL for event-triggered control | Optimized setting and event-trigged control | Occupants' discomfort rate: proportional to the square of deviation of desired temperature and the coefficient of proportionality $r_c{}^2$ | Indoor/Outdoor temperature; Heater status (0 when off/1 when on); Desired temperature | Simulation results showed that the proposed algorithms learn the optimal policy in an appropriate time, i.e., optimal thresholds were found $T_{ON}^{th} = 12.5°C$ and $T_{OFF}^{th} = 17.5°C$ resulting an optimal rewards rate. | [149] |
| 2020 | A neural network-based approach for energy management and climate control optimization of buildings (applied to two-story building in Italy). | Historical data (October 2018-February 2019, and October-November 2019) from the building automation system and a weather station provided by the CETEMPS (http://cetemps.aquila.infn.it) | MPC with neural network-based models | Predictive control and optimized setting | Constant set-point temperature (defined as $T_{ref} = 25°C$) for each zone. | **Control inputs:** Temperature set-points; Compressor mode (Boolean) **Disturbances:** Outside temperature; Humidity; Solar radiation | The proposed model showed significant results in energy savings (5.7% energy reduction of one zone) and better comfort compared to the baseline controller. | [150] |
| 2020 | Comfort and energy management of daily and seasonally used appliances for smart buildings application in hottest areas. | Data, including appliances and power rating as well as the occupants' data, were taken from the reference [151] | Binary Particle Swarm Optimization (BPSO) + Fuzzy logic: Mamdani FIS & Sugeno FIS | Two proposed controllers: BPSOFMAM and BPSOFSUG | Fanger's PMV method | Room temperature; Outdoor temperature; Initialized set-points; Occupancy level; Price | Simulation results showed that the BPSOFSUG controller outperformed the BPSOFMAM in terms of energy efficiency by 16%, while comfort computation, via PMV, was kept in satisfactory range. | [152] |
| 2020 | Thermal comfort control relying on a smart WiFi-based thermostat deployment for residential applications | Data collected from 700+ university student residences in the Midwest USA: historical WiFi thermostat readings, monthly energy consumption, building geometry, and weather data obtained from NOAA's Climate Data Online resource [3] | Nonlinear autoregressive network with exogenous inputs (NARX) neural network using Levenberg–Marquart as training function | Learning-based predictive control | Fanger's PMV method | Building Geometry; Occupancy level; Comfort parameters[4]; Human times; Cool/heat/fan status; Cooling/heating set-point; Indoor air temperature; Relative humidity; Outdoor weather conditions | The proposed dynamic model has showed in both high- and low-efficiency residences, cooling energy savings were around 85% and 95% respectively, while the PMV index was maintained within the desired rang [0 – 0.5]. | [153] |
| 2020 | A network-based deterministic model development to respond the ever-changing users' fickle taste that can deteriorate thermal comfort and energy efficiency in building spaces. | Data obtained from actual devices installed in buildings/Questionnaire surveys of users. | Fuzzy inference system (FIS) to determine abnormal situations + ANN | Three controllers were tested: Thermostat On/Off; ANN; ANN + FDM | Fanger's PMV method | Heat loss of the room (HL); Relative humidity (RH); Metabolic rate (MET); Clothing insulation (Clo) | ANN-FDM showed significant results by improving thermal comfort by up to 4.3% rather than thermostat model and up to 44.1% of energy efficiency rather than ANN model. | [154] |

---

[2] $r_c = -1.2/3600 \; unit \; K^{-2}s^{-1}$
[3] https://catalog.data.gov/dataset/noaas-climate-divisional-database-nclimdiv
[4] Comfort parameters: room air velocity, clothing level, metabolic rate, mean radiant temperature, room temperature and relative humidity.

**Table 1.** Cont.

| Year | Study Case | Source of Data used | Underlying AI/ML techniques | Application Scenario | Thermal comfort-based model/ method | Input(s) & Controlled parameter(s) | Key Results | Ref |
|---|---|---|---|---|---|---|---|---|
| 2020 | ANN-based prognostic models' development for load demand (LD) prediction for a Greek island by capturing three different forecasting horizons: medium, short and very short-terms. | Meteorological and LD data collected from the island of Tilos in Greece covering a period from April 2015 to April 2017. | Multilayer Perception ANN and stochastic/persistence autoregressive (AR) time series forecasting models' development for load demand prediction. | Predictive control | Biometeorological human thermal comfort-discomfort index: Cooling power (CP) index $CP = 1.163 \cdot (10.45 + 10 \cdot u^{0.5} - u) \cdot (33 - T)$ {T: temperature, °C and u: wind speed, m/s} | Relative humidity; Barometric pressure; Solar irradiation; Cooling Power index = {Air temperature, Wind speed} | Results showed that both medium- and short-terms prognoses shown significant ability to predict LD by errors around 7.9% and 7.2% respectively enabling a better management for end-user and energy. | [155] |
| 2020 | An intelligent-based ML model to predict the energy performances in heating loads (HL) and cooling loads (CL) of residential buildings. | Dataset freely available at the Center of machine learning and intelligent systems repository [5], where 768 buildings located in Athens, Greece were simulated using Ecotect software. | ANN and Deep NN models were evaluated for CL and HL forecasting. | Predictive control | Comfort conditions considered in the internal design of the buildings, i.e., clothing level of 0.6 Clo with internal temperature of 21°C, 60% of humidity, 0.3 m/s air speed and 300 Lux lighting level | Relative Compactness; Surface Area; Wall Area; Roof Area; Overall Height; Orientation; Glazing Area; Glazing Area Distribution | Deep NN showed better results compared to ANN in terms of HL and CL prediction, by applying state-based sensitivity analysis (SBSA) technique allowing to improve the model by selecting the most significant variables. | [156] |
| 2020 | A novel personal thermal comfort prediction method using less physiological parameters. | 45 experiments were conducted with 3 subjects in an office room in Shanghai, thermal sensation surveys/ questionnaires of the occupants to collect personal information, physiological parameters + environmental variables measurements | ANN was used to evaluate the thermal sensation | Predictive control | Thermal sensation vote classified into 5 categories {cold, cool, neutral, warm, hot} | **Personal information:** Name; Sex; Age; Height; Weight; BMI; $I_{cl}$; Thermal sensation vote (TSV) **Environmental parameters:** Air temperature; Air humidity **Physiological parameters:** Skin temperatures of the wrist/the neck; temperature of the point 2 mm above the wrist | Based on the 3 physiological parameters, the proposed model showed good prediction accuracy and stability by an average of 89.2% and a standard deviation around 2.0%, this model will be used in HVAC operations for energy savings as well. | [157] |
| 2020 | Investigating the performances and comparative analyses of combined on-demand and predictive models for thermal conditions control in buildings. | Geometries/Design parameters of the building's model and simulation parameters adopted from templates[101], [102], [158] and ASHRAE9012016_School Secondary in the EnergyPlus. Weather data file obtained from EnergyPlus Weather Data website. | Combining ANN and the fuzzy inference system (FIS) to control supply air mass and its temperature | On-demand and predictive controls | Fanger's PMV/PPD method | Climate conditions; Building geometry; Design parameters; Indoor temperature; Human Comfort (PMV/PPD); Outdoor temperature | The combination of the predictive and on-demand algorithms improved the energy efficiency from 13.1% to 44.4% and reduced the thermal dissatisfaction by 20% to 33.6%, compared to each independent model. | [159] |
| 2020 | A multi-objective optimization method for a passive house (PH) design by considering energy demand, thermal comfort and cost. | Simulation data using EnergyPlus, weather data of Tianjin city (case location) derived from the Chinese Standard Weather Data published by EnergyPlus Website | Three methods were combined: Redundancy analysis (RDA), Gradient Boosted Decision Trees (GBDT) and Non-dominated sorting GA (NSGA-II) for multi-optimization purpose | Optimized setting | The annual cumulative comfort ratio (CTR)-based adaptive model $CTR = \frac{1}{m}\sum_{k=1}^{k=m}\left(\sum_{j=1}^{N_p} wf_j \cdot \frac{1}{N_p}\right)^m \in [0,1]$ | Building features: Wall and roof; Exterior windows; Building shape; Airtightness and building layout | The implemented model outperformed other tested methods (SVM and ANN) with a SD=0.048. the optimization results showed around 88.2% energy savings rate and improvement in thermal comfort by 37.7% compared to base-case building | [160] |
| 2020 | A predictive model for thermal energy by integrating IoT architecture based on *Edge Computing* and classifier ensemble techniques for smart buildings application. | Recorded real-sensor data through a monitoring-system including parameters of buildings multiplicity, corresponding to one-month in the year of 2014. | Combining classifier techniques: SVM, logistic regression (LR) and random forest (RF) for temperature prediction in conditioned spaces | Predictive control | Indoor temperature was set by the user or by the learning algorithm (by considering the user and defined temperature by the proposed methodology) | Temperature of dining room; $CO_2$ of dining room; Temperature of indoor room; Relative humidity of Dining room/Room; Room lighting; Solar irradiance; Day of week | Simulation results showed that the proposed approach presented the highest accuracy, by 91.526% compared to neural networks, ensemble RF and SVM. | [161] |

---

[5] http://archive.ics.uci.edu/ml/datasets.php

## 4. Final Considerations and Conclusion

In this paper, 120 published works related to BEMS and EMS for equipment control while considering the comfort factor using the artificial intelligence techniques have been investigated. From 1993 to 2020, the application of AI-based tools were analyzed in order to evaluate their performances in energy savings and thermal comfort optimization. Statistical results showed that the energy savings, on average, is between 21.81% and 44.36%, while thermal comfort enhancement is between 19.75% and 85.77%, on average as well. On the other hand, we have proved that there is a direct relationship between the energy consumption and the way or methods used to infer thermal comfort. Indeed, based on the statistical results, it can be observed that the highest comfort levels come from tools integrating the human factor into the loop. For example, GA methods could reach an average comfort improvement of 85.77% by adopting the user preferences or the Fanger's model.

## References

[1]  O. de coopération et de développement économiques, *Transition to sustainable buildings: Strategies and opportunities to 2050*. OECD Publishing, 2013.
[2]  E. W. Shaw, "Thermal Comfort: analysis and applications in environmental engineering, by P. O. Fanger. 244 pp. DANISH TECHNICAL PRESS. Copenhagen, Denmark, 1970. Danish Kr. 76, 50," vol. 92, Jun. 1972, Accessed: May 17, 2019. [Online]. Available: https://doi.org/10.1177/146642407209200337.
[3]  M. Sherman, "A simplified model of thermal comfort," *Energy and Buildings*, vol. 8, no. 1, pp. 37–50, Feb. 1985, doi: 10.1016/0378-7788(85)90013-1.
[4]  A. P. Gagge, A. P. Fobelets, and L. G. Berglund, "A standard predictive index of human response to the thermal environment," Research Org.: John B. Pierce Foundation Lab., New Haven, CT, Jan. 1986, [Online]. Available: https://www.osti.gov/servlets/purl/6494216.
[5]  R. and A.-C. Engineers. American Society of Heating, *1993 ASHRAE handbook : fundamentals*. Atlanta, GA.: ASHRAE, 1993.
[6]  F. Oldewurtel *et al.*, "Increasing Energy Efficiency in Building Climate Control using Weather Forecasts and Model Predictive Control," p. 8.
[7]  M. Gwerder *et al.*, "Potential Assessment of Rule-Based Control for Integrated Room Automation," in *10th REHVA World Congress, "Sustainable Energy Use in Buildings" – Clima 2010*, Antalya, Turquia, 2010, p. 8.
[8]  A. I. Dounis and C. Caraiscos, "Advanced control systems engineering for energy and comfort management in a building environment—A review," *Renewable and Sustainable Energy Reviews*, vol. 13, no. 6, pp. 1246–1261, 2009, doi: https://doi.org/10.1016/j.rser.2008.09.015.
[9]  G. Oliveira, L. S. COELHO, N. Mendes, H. X. Araujo, and others, "TED-AJ03-258 Using Fuzzy Logic in Heating Control Systems," in *Proceedings of the... ASME/JSME Thermal Engineering Joint Conference*, 2003, vol. 2003, p. 74.
[10] M. M. Gouda, S. Danaher, and C. P. Underwood, "Thermal comfort based fuzzy logic controller," *Building Services Engineering Research and Technology*, vol. 22, no. 4, pp. 237–253, Nov. 2001, doi: 10.1177/014362440102200403.
[11] A. I. Dounis and C. Caraiscos, "Intelligent Coordinator of Fuzzy Controller-Agents for Indoor Environment Control in Buildings Using 3-D Fuzzy Comfort Set," in *2007 IEEE International Fuzzy Systems Conference*, London, UK, Jun. 2007, pp. 1–6, doi: 10.1109/FUZZY.2007.4295573.
[12] Z. Yu, Y. Zhou, and A. Dexter, "Hierarchical fuzzy rule-based control of renewable energy building systems," p. 6, Jan. 2007.
[13] F. Calvino, M. La Gennusa, G. Rizzo, and G. Scaccianoce, "The control of indoor thermal comfort conditions: introducing a fuzzy adaptive controller," *Energy and Buildings*, vol. 36, no. 2, pp. 97–102, Feb. 2004, doi: 10.1016/j.enbuild.2003.10.004.
[14] Y. Yan, J. Zhou, Y. Lin, W. Yang, P. Wang, and G. Zhang, "Adaptive optimal control model for building cooling and heating sources," *Energy and Buildings*, vol. 40, no. 8, pp. 1394–1401, Jan. 2008, doi: 10.1016/j.enbuild.2008.01.003.
[15] S. Atthajariyakul and T. Leephakpreeda, "Real-time determination of optimal indoor-air condition for thermal comfort, air quality and efficient energy usage," *Energy and Buildings*, vol. 36, no. 7, pp. 720–733, Jul. 2004, doi: 10.1016/j.enbuild.2004.01.017.
[16] W. Liu, Z. Lian, and B. Zhao, "A neural network evaluation model for individual thermal comfort," *Energy and Buildings*, vol. 39, no. 10, pp. 1115–1122, Oct. 2007, doi: 10.1016/j.enbuild.2006.12.005.

[17]   Jian Liang and Ruxu Du, "Thermal comfort control based on neural network for HVAC application," in *Proceedings of 2005 IEEE Conference on Control Applications, 2005. CCA 2005.*, Toronto, Canada, 2005, pp. 819–824, doi: 10.1109/CCA.2005.1507230.
[18]   M. Hadjiski, N. Deliiski, and K. Boshnakov, "Thermal Comfort-based MPC of Air Handling Unit," *IFAC Proceedings Volumes*, vol. 39, no. 19, pp. 81–86, Jan. 2006, doi: 10.3182/20061002-4-BG-4905.00014.
[19]   E. Donaisky, G. H. C. Oliveira, R. Z. Freire, and N. Mendes, "PMV-Based Predictive Algorithms for Controlling Thermal Comfort in Building Plants," in *2007 IEEE International Conference on Control Applications*, Oct. 2007, pp. 182–187, doi: 10.1109/CCA.2007.4389227.
[20]   D. Moher, A. Liberati, J. Tetzlaff, D. G. Altman, and PRISMA Group, "Preferred reporting items for systematic reviews and meta-analyses: the PRISMA statement," *PLoS Med*, vol. 6, no. 7, pp. 336–341, Jul. 2009, doi: 10.1371/journal.pmed.1000097.
[21]   A. Liberati *et al.*, "The PRISMA statement for reporting systematic reviews and meta-analyses of studies that evaluate health care interventions: explanation and elaboration," *Annals of internal medicine*, vol. 151, no. 4, p. W–65, 2009, doi: 10.1371/journal.pmed.1000100.
[22]   H. Zhou, M. Rao, and K. T. Chuang, "Knowledge-based automation for energy conservation and indoor air quality control in HVAC processes," *Engineering Applications of Artificial Intelligence*, vol. 6, no. 2, pp. 131–144, Apr. 1993, doi: 10.1016/0952-1976(93)90029-W.
[23]   M. Hamdi and G. Lachiver, "A fuzzy control system based on the human sensation of thermal comfort," in *1998 IEEE International Conference on Fuzzy Systems Proceedings. IEEE World Congress on Computational Intelligence (Cat. No.98CH36228)*, Anchorage, AK, USA, 1998, vol. 1, pp. 487–492, doi: 10.1109/FUZZY.1998.687534.
[24]   Th. Bernard and H.-B. Kuntze, "Multi-objective optimization of building climate control systems using fuzzy-logic," in *1999 European Control Conference (ECC)*, Karlsruhe, Aug. 1999, pp. 2572–2577, doi: 10.23919/ECC.1999.7099712.
[25]   D. Kolokotsa, D. Tsiavos, G. S. Stavrakakis, K. Kalaitzakis, and E. Antonidakis, "Advanced fuzzy logic controllers design and evaluation for buildings' occupants thermal–visual comfort and indoor air quality satisfaction," *Energy and Buildings*, vol. 33, no. 6, pp. 531–543, Jul. 2001, doi: 10.1016/S0378-7788(00)00098-0.
[26]   D. Kolokotsa, G. S. Stavrakakis, K. Kalaitzakis, and D. Agoris, "Genetic algorithms optimized fuzzy controller for the indoor environmental management in buildings implemented using PLC and local operating networks," *Engineering Applications of Artificial Intelligence*, vol. 15, no. 5, pp. 417–428, Sep. 2002, doi: 10.1016/S0952-1976(02)00090-8.
[27]   R. Alcalá, J. M. Benítez, J. Casillas, J. Casillas, O. Cordón, and R. Pérez, "Fuzzy Control of HVAC Systems Optimized by Genetic Algorithms," *Applied Intelligence*, vol. 18, no. 2, pp. 155–177, 2003.
[28]   D. Kolokotsa, "Comparison of the performance of fuzzy controllers for the management of the indoor environment," *Building and Environment*, vol. 38, no. 12, pp. 1439–1450, Dec. 2003, doi: 10.1016/S0360-1323(03)00130-6.
[29]   A. B. Shepherd and W. J. Batty, "Fuzzy control strategies to provide cost and energy efficient high quality indoor environments in buildings with high occupant densities," *Building Services Engineering Research and Technology*, vol. 24, no. 1, pp. 35–45, Feb. 2003, doi: 10.1191/0143624403bt059oa.
[30]   N. Nassif, S. Kajl, and R. Sabourin, "Two-objective on-line optimization of supervisory control strategy," *Building Services Engineering Research and Technology*, vol. 25, no. 3, pp. 241–251, Aug. 2004, doi: 10.1191/0143624404bt105oa.
[31]   R. Alcalá, J. Casillas, O. Cordón, A. González, and F. Herrera, "A genetic rule weighting and selection process for fuzzy control of heating, ventilating and air conditioning systems," *Engineering Applications of Artificial Intelligence*, vol. 18, no. 3, pp. 279–296, Apr. 2005, doi: 10.1016/j.engappai.2004.09.007.
[32]   S. Ari, I. A. Cosden, H. E. Khalifa, J. F. Dannenhoffer, P. Wilcoxen, and C. Isik, "Constrained Fuzzy Logic Approximation for Indoor Comfort and Energy Optimization," in *NAFIPS 2005 - 2005 Annual Meeting of the North American Fuzzy Information Processing Society*, Detroit, MI, USA, 2005, pp. 500–504, doi: 10.1109/NAFIPS.2005.1548586.
[33]   P. Davidsson and M. Boman, "Distributed monitoring and control of office buildings by embedded agents," *Information Sciences*, vol. 171, no. 4, pp. 293–307, May 2005, doi: 10.1016/j.ins.2004.09.007.
[34]   D. Kolokotsa, K. Kalaitzakis, E. Antonidakis, and G. Stavrakakis, "Interconnecting smart card system with PLC controller in a local operating network to form a distributed energy management and control system for buildings," *Energy conversion and Management*, vol. 43, no. 1, pp. 119–134, 2002.
[35]   D. Kolokotsa, K. Niachou, V. Geros, K. Kalaitzakis, G. S. Stavrakakis, and M. Santamouris, "Implementation of an integrated indoor environment and energy management system," *Energy and Buildings*, vol. 37, no. 1, pp. 93–99, Jan. 2005, doi: 10.1016/j.enbuild.2004.05.008.
[36]   M. T. Lah, B. Zupančič, and A. Krainer, "Fuzzy control for the illumination and temperature comfort in a test chamber," *Building and Environment*, vol. 40, no. 12, pp. 1626–1637, Dec. 2005, doi: 10.1016/j.buildenv.2004.11.008.

[37] M. Hadjiski, V. Sgurev, and V. Boishina, "Multi-Agent Intelligent Control of Centralized HVAC Systems," *IFAC Proceedings Volumes*, vol. 39, no. 19, pp. 195–200, 2006, doi: 10.3182/20061002-4-BG-4905.00033.

[38] Y. Huang and N. Li, "Indoor Thermal Comfort Control Research Based on Adaptive Fuzzy Strategy," in *Computational Engineering in Systems Applications*, 2006, p. 4.

[39] K. Dalamagkidis, D. Kolokotsa, K. Kalaitzakis, and G. S. Stavrakakis, "Reinforcement learning for energy conservation and comfort in buildings," *Building and Environment*, vol. 42, no. 7, pp. 2686–2698, Jul. 2007, doi: 10.1016/j.buildenv.2006.07.010.

[40] G. J. Ríos-Moreno, M. Trejo-Perea, R. Castañeda-Miranda, V. M. Hernández-Guzmán, and G. Herrera-Ruiz, "Modelling temperature in intelligent buildings by means of autoregressive models," *Automation in Construction*, vol. 16, no. 5, pp. 713–722, Aug. 2007, doi: 10.1016/j.autcon.2006.11.003.

[41] E. Sierra, A. Hossian, P. Britos, D. Rodriguez, and R. Garcia-Martinez, "Fuzzy Control for Improving Energy Management within Indoor Building Environments," in *Electronics, Robotics and Automotive Mechanics Conference (CERMA 2007)*, Cuernavaca, Morelos, Mexico, Sep. 2007, pp. 412–416, doi: 10.1109/CERMA.2007.4367722.

[42] J. Liang and R. Du, "Design of intelligent comfort control system with human learning and minimum power control strategies," *Energy Conversion and Management*, vol. 49, no. 4, pp. 517–528, Apr. 2008, doi: 10.1016/j.enconman.2007.08.006.

[43] S. Jassar, Z. Liao, and L. Zhao, "Adaptive neuro-fuzzy based inferential sensor model for estimating the average air temperature in space heating systems," *Building and Environment*, vol. 44, no. 8, pp. 1609–1616, Aug. 2009, doi: 10.1016/j.buildenv.2008.10.002.

[44] M. Mossolly, K. Ghali, and N. Ghaddar, "Optimal control strategy for a multi-zone air conditioning system using a genetic algorithm," *Energy*, vol. 34, no. 1, pp. 58–66, Jan. 2009, doi: 10.1016/j.energy.2008.10.001.

[45] S. Soyguder and H. Alli, "Predicting of fan speed for energy saving in HVAC system based on adaptive network based fuzzy inference system," *Expert Systems with Applications*, vol. 36, no. 4, pp. 8631–8638, May 2009, doi: 10.1016/j.eswa.2008.10.033.

[46] R. J. De Dear, "A global database of thermal comfort field experiments," *ASHRAE transactions*, vol. 104, p. 1141, 1998.

[47] J. Toftum, R. V. Andersen, and K. L. Jensen, "Occupant performance and building energy consumption with different philosophies of determining acceptable thermal conditions," *Building and Environment*, vol. 44, no. 10, pp. 2009–2016, Oct. 2009, doi: 10.1016/j.buildenv.2009.02.007.

[48] Y. Gao, E. Tumwesigye, B. Cahill, and K. Menzel, "Using Data Mining in Optimisation of Building Energy Consumption and Thermal Comfort Management," in *The 2nd International Conference on Software Engineering and Data Mining*, Chengdu, China, Jun. 2010, pp. 434–439.

[49] L. Magnier and F. Haghighat, "Multiobjective optimization of building design using TRNSYS simulations, genetic algorithm, and Artificial Neural Network," *Building and Environment*, vol. 45, no. 3, pp. 739–746, Mar. 2010, doi: 10.1016/j.buildenv.2009.08.016.

[50] A. I. Dounis, P. Tiropanis, A. Argiriou, and A. Diamantis, "Intelligent control system for reconciliation of the energy savings with comfort in buildings using soft computing techniques," *Energy and Buildings*, vol. 43, no. 1, pp. 66–74, Jan. 2011, doi: 10.1016/j.enbuild.2010.08.014.

[51] G. Jahedi and M. M. Ardehali, "Genetic algorithm-based fuzzy-PID control methodologies for enhancement of energy efficiency of a dynamic energy system," *Energy Conversion and Management*, vol. 52, no. 1, pp. 725–732, Jan. 2011, doi: 10.1016/j.enconman.2010.07.051.

[52] L. Klein *et al.*, "Towards Optimization of Building Energy and Occupant Comfort Using Multi-Agent Simulation," presented at the 28th International Symposium on Automation and Robotics in Construction, Seoul, Korea, Jun. 2011, doi: 10.22260/ISARC2011/0044.

[53] U. C. Bureau, "American Housing Survey (AHS)," *The United States Census Bureau*. https://www.census.gov/programs-surveys/ahs.html (accessed Mar. 11, 2020).

[54] J. W. Moon, S. K. Jung, Y. Kim, and S.-H. Han, "Comparative study of artificial intelligence-based building thermal control methods – Application of fuzzy, adaptive neuro-fuzzy inference system, and artificial neural network," *Applied Thermal Engineering*, vol. 31, no. 14–15, pp. 2422–2429, Oct. 2011, doi: 10.1016/j.applthermaleng.2011.04.006.

[55] "Local Weather Forecast, News and Conditions | Weather Underground." https://www.wunderground.com/ (accessed Mar. 11, 2020).

[56] Z. Wang, R. Yang, L. Wang, R. C. Green, and A. I. Dounis, "A fuzzy adaptive comfort temperature model with grey predictor for multi-agent control system of smart building," in *2011 IEEE Congress of Evolutionary Computation (CEC)*, New Orleans, LA, USA, Jun. 2011, pp. 728–735, doi: 10.1109/CEC.2011.5949691.

[57] R. Yang and L. Wang, "Energy management of multi-zone buildings based on multi-agent control and particle swarm optimization," in *2011 IEEE International Conference on Systems, Man, and Cybernetics*, Anchorage, AK, USA, Oct. 2011, pp. 159–164, doi: 10.1109/ICSMC.2011.6083659.

[58] A. Aswani, N. Master, J. Taneja, D. Culler, and C. Tomlin, "Reducing Transient and Steady State Electricity Consumption in HVAC Using Learning-Based Model-Predictive Control," *Proceedings of the IEEE*, vol. 100, no. 1, pp. 240–253, Jan. 2012, doi: 10.1109/JPROC.2011.2161242.

[59] F. A. Barata and R. Neves-Silva, "Distributed model predictive control for thermal house comfort with auction of available energy," in *2012 International Conference on Smart Grid Technology, Economics and Policies (SG-TEP)*, Nuremberg, Germany, Dec. 2012, pp. 1–4, doi: 10.1109/SG-TEP.2012.6642375.

[60] P. M. Ferreira, A. E. Ruano, S. Silva, and E. Z. E. Conceição, "Neural networks based predictive control for thermal comfort and energy savings in public buildings," *Energy and Buildings*, vol. 55, pp. 238–251, Dec. 2012, doi: 10.1016/j.enbuild.2012.08.002.

[61] L. Klein *et al.*, "Coordinating occupant behavior for building energy and comfort management using multi-agent systems," *Automation in Construction*, vol. 22, pp. 525–536, Mar. 2012, doi: 10.1016/j.autcon.2011.11.012.

[62] Z. Wang, L. Wang, A. I. Dounis, and R. Yang, "Multi-agent control system with information fusion based comfort model for smart buildings," *Applied Energy*, vol. 99, pp. 247–254, Nov. 2012, doi: 10.1016/j.apenergy.2012.05.020.

[63] P. X. Gao and S. Keshav, "Optimal Personal Comfort Management Using SPOT+," in *Proceedings of the 5th ACM Workshop on Embedded Systems For Energy-Efficient Buildings - BuildSys'13*, Roma, Italy, 2013, pp. 1–8, doi: 10.1145/2528282.2528297.

[64] C. Li, G. Zhang, M. Wang, and J. Yi, "Data-driven modeling and optimization of thermal comfort and energy consumption using type-2 fuzzy method," *Soft Computing*, vol. 17, no. 11, pp. 2075–2088, Nov. 2013, doi: 10.1007/s00500-013-1117-4.

[65] S. D. Smitha, J. S. Savier, and F. Mary Chacko, "Intelligent control system for efficient energy management in commercial buildings," in *2013 Annual International Conference on Emerging Research Areas and 2013 International Conference on Microelectronics, Communications and Renewable Energy*, Kanjirapally, India, Jun. 2013, pp. 1–6, doi: 10.1109/AICERA-ICMiCR.2013.6575942.

[66] D. Wijayasekara, M. Manic, and C. Rieger, "Fuzzy linguistic knowledge based behavior extraction for building energy management systems," in *2013 6th International Symposium on Resilient Control Systems (ISRCS)*, San Francisco, CA, USA, Aug. 2013, pp. 80–85, doi: 10.1109/ISRCS.2013.6623755.

[67] R. Yang and L. Wang, "Development of multi-agent system for building energy and comfort management based on occupant behaviors," *Energy and Buildings*, vol. 56, pp. 1–7, Jan. 2013, doi: 10.1016/j.enbuild.2012.10.025.

[68] M. Collotta, A. Messineo, G. Nicolosi, and G. Pau, "A Dynamic Fuzzy Controller to Meet Thermal Comfort by Using Neural Network Forecasted Parameters as the Input," *Energies*, vol. 7, no. 8, pp. 4727–4756, Jul. 2014, doi: 10.3390/en7084727.

[69] B. Dong and K. P. Lam, "A real-time model predictive control for building heating and cooling systems based on the occupancy behavior pattern detection and local weather forecasting," *Building Simulation*, vol. 7, no. 1, pp. 89–106, Feb. 2014, doi: 10.1007/s12273-013-0142-7.

[70] P. Fazenda, K. Veeramachaneni, P. Lima, and U.-M. O'Reilly, "Using reinforcement learning to optimize occupant comfort and energy usage in HVAC systems," *JAISE*, vol. 6, no. 6, pp. 675–690, Nov. 2014, doi: 10.3233/AIS-140288.

[71] A. Ghahramani, F. Jazizadeh, and B. Becerik-Gerber, "A knowledge based approach for selecting energy-aware and comfort-driven HVAC temperature set points," *Energy and Buildings*, vol. 85, pp. 536–548, Dec. 2014, doi: 10.1016/j.enbuild.2014.09.055.

[72] S. Hussain, H. A. Gabbar, D. Bondarenko, F. Musharavati, and S. Pokharel, "Comfort-based fuzzy control optimization for energy conservation in HVAC systems," *Control Engineering Practice*, vol. 32, pp. 172–182, Nov. 2014, doi: 10.1016/j.conengprac.2014.08.007.

[73] A. Javed, H. Larijani, A. Ahmadinia, and R. Emmanuel, "Modelling and optimization of residential heating system using random neural networks," in *2014 IEEE International Conference on Control Science and Systems Engineering*, Yantai, China, Dec. 2014, pp. 90–95, doi: 10.1109/CCSSE.2014.7224515.

[74] F. Jazizadeh, A. Ghahramani, B. Becerik-Gerber, T. Kichkaylo, and M. Orosz, "User-led decentralized thermal comfort driven HVAC operations for improved efficiency in office buildings," *Energy and Buildings*, vol. 70, pp. 398–410, Feb. 2014, doi: 10.1016/j.enbuild.2013.11.066.

[75] J. Langevin, J. Wen, and P. L. Gurian, "Including Occupants in Building Performance Simulation: Integration of an Agent-Based Occupant Behavior Algorithm with EnergyPlus," in *2014 ASHRAE/IBPSA-USA Buidling Simulation Conference. Atlanta, GA*, Atlanta, GA, USA, 2014, p. 8.

[76] M. Mokhtar, X. Liu, and J. Howe, "Multi-agent Gaussian Adaptive Resonance Theory Map for building energy control and thermal comfort management of UCLan's WestLakes Samuel Lindow Building," *Energy and Buildings*, vol. 80, pp. 504–516, Sep. 2014, doi: 10.1016/j.enbuild.2014.05.045.

[77] J. W. Moon, J.-H. Lee, and S. Kim, "Application of control logic for optimum indoor thermal environment in buildings with double skin envelope systems," *Energy and Buildings*, vol. 85, pp. 59–71, Dec. 2014, doi: 10.1016/j.enbuild.2014.09.018.

[78] P. H. Shaikh, N. B. M. Nor, P. Nallagownden, and I. Elamvazuthi, "Stochastic optimized intelligent controller for smart energy efficient buildings," *Sustainable Cities and Society*, vol. 13, pp. 41–45, Oct. 2014, doi: 10.1016/j.scs.2014.04.005.

[79] R. Emmanuel, C. Clark, A. Ahmadinia, A. Javed, D. Gibson, and H. Larijani, "Experimental testing of a random neural network smart controller using a single zone test chamber," *IET Networks*, vol. 4, no. 6, pp. 350–358, Nov. 2015, doi: 10.1049/iet-net.2015.0020.

[80] A. Garnier, J. Eynard, M. Caussanel, and S. Grieu, "Predictive control of multizone heating, ventilation and air-conditioning systems in non-residential buildings," *Applied Soft Computing*, vol. 37, pp. 847–862, Dec. 2015, doi: 10.1016/j.asoc.2015.09.022.

[81] J. W. Moon, "Comparative performance analysis of the artificial-intelligence-based thermal control algorithms for the double-skin building," *Applied Thermal Engineering*, vol. 91, pp. 334–344, Dec. 2015, doi: 10.1016/j.applthermaleng.2015.08.038.

[82] Hao Huang, Lei Chen, and E. Hu, "A hybrid model predictive control scheme for energy and cost savings in commercial buildings: Simulation and experiment," in *2015 American Control Conference (ACC)*, Chicago, IL, USA, Jul. 2015, pp. 256–261, doi: 10.1109/ACC.2015.7170745.

[83] L. A. Hurtado, P. H. Nguyen, and W. L. Kling, "Smart grid and smart building inter-operation using agent-based particle swarm optimization," *Sustainable Energy, Grids and Networks*, vol. 2, pp. 32–40, Jun. 2015, doi: 10.1016/j.segan.2015.03.003.

[84] C.-S. Kang, C.-H. Hyun, and M. Park, "Fuzzy logic-based advanced on–off control for thermal comfort in residential buildings," *Applied Energy*, vol. 155, pp. 270–283, Oct. 2015, doi: 10.1016/j.apenergy.2015.05.119.

[85] K. L. Ku, J. S. Liaw, M. Y. Tsai, and T. S. Liu, "Automatic Control System for Thermal Comfort Based on Predicted Mean Vote and Energy Saving," *IEEE Transactions on Automation Science and Engineering*, vol. 12, no. 1, pp. 378–383, Jan. 2015, doi: 10.1109/TASE.2014.2366206.

[86] B. Li and L. Xia, "A multi-grid reinforcement learning method for energy conservation and comfort of HVAC in buildings," in *2015 IEEE International Conference on Automation Science and Engineering (CASE)*, Gothenburg, Sweden, Aug. 2015, pp. 444–449, doi: 10.1109/CoASE.2015.7294119.

[87] D. Zupančič, M. Luštrek, and M. Gams, "Multi-Agent Architecture for Control of Heating and Cooling in a Residential Space," *The Computer Journal*, vol. 58, no. 6, pp. 1314–1329, Jun. 2015, doi: 10.1093/comjnl/bxu058.

[88] ASHRAE, *International Weather for Energy Calculations (IWEC Weather Files) Users Manual and CD-ROM*. ASHRAE Atlanta, 2001.

[89] F. Ascione, N. Bianco, C. De Stasio, G. M. Mauro, and G. P. Vanoli, "Simulation-based model predictive control by the multi-objective optimization of building energy performance and thermal comfort," *Energy and Buildings*, vol. 111, pp. 131–144, Jan. 2016, doi: 10.1016/j.enbuild.2015.11.033.

[90] K. Dornelles, V. Roriz, and M. Roriz, "Determination of the solar absorptance of opaque surfaces," in *24th International Conference on Passive and Low Energy Architecture*, 2007, pp. 452–9.

[91] N. Delgarm, B. Sajadi, and S. Delgarm, "Multi-objective optimization of building energy performance and indoor thermal comfort: A new method using artificial bee colony (ABC)," *Energy and Buildings*, vol. 131, pp. 42–53, Nov. 2016, doi: 10.1016/j.enbuild.2016.09.003.

[92] X. Li, J. Wen, and E.-W. Bai, "Developing a whole building cooling energy forecasting model for on-line operation optimization using proactive system identification," *Applied Energy*, vol. 164, pp. 69–88, 2016.

[93] X. Li, J. Wen, and A. Malkawi, "An operation optimization and decision framework for a building cluster with distributed energy systems," *Applied Energy*, vol. 178, pp. 98–109, Sep. 2016, doi: 10.1016/j.apenergy.2016.06.030.

[94] J. W. Moon and S. K. Jung, "Algorithm for optimal application of the setback moment in the heating season using an artificial neural network model," *Energy and Buildings*, vol. 127, pp. 859–869, Sep. 2016, doi: 10.1016/j.enbuild.2016.06.046.

[95] J. W. Moon and S. K. Jung, "Development of a thermal control algorithm using artificial neural network models for improved thermal comfort and energy efficiency in accommodation buildings," *Applied Thermal Engineering*, vol. 103, pp. 1135–1144, Jun. 2016, doi: 10.1016/j.applthermaleng.2016.05.002.

[96] P. H. Shaikh, N. B. M. Nor, P. Nallagownden, I. Elamvazuthi, and T. Ibrahim, "Intelligent multi-objective control and management for smart energy efficient buildings," *International Journal of Electrical Power & Energy Systems*, vol. 74, pp. 403–409, Jan. 2016, doi: 10.1016/j.ijepes.2015.08.006.

[97]   M. Rasheed *et al.*, "Real Time Information Based Energy Management Using Customer Preferences and Dynamic Pricing in Smart Homes," *Energies*, vol. 9, no. 7, p. 542, Jul. 2016, doi: 10.3390/en9070542.

[98]   G. Mestre *et al.*, "An Intelligent Weather Station," *Sensors*, vol. 15, no. 12, pp. 31005–31022, 2015, doi: 10.3390/s151229841.

[99]   A. E. Ruano *et al.*, "The IMBPC HVAC system: A complete MBPC solution for existing HVAC systems," *Energy and Buildings*, vol. 120, pp. 145–158, May 2016, doi: 10.1016/j.enbuild.2016.03.043.

[100]  M. Soudari, S. Srinivasan, S. Balasubramanian, J. Vain, and U. Kotta, "Learning based personalized energy management systems for residential buildings," *Energy and Buildings*, vol. 127, pp. 953–968, Sep. 2016, doi: 10.1016/j.enbuild.2016.05.059.

[101]  A. Ashrae, "Standard 55-2004," *Thermal environmental conditions for human occupancy*, vol. 744, 2004.

[102]  "Indoor Design Conditions - Summer and Winter," *Engineering Toolbox*. https://www.engineeringtoolbox.com/inside-design-conditions-d_1570.html (accessed Mar. 15, 2020).

[103]  J. Ahn and S. Cho, "Anti-logic or common sense that can hinder machine's energy performance: Energy and comfort control models based on artificial intelligence responding to abnormal indoor environments," *Applied Energy*, vol. 204, pp. 117–130, Oct. 2017, doi: 10.1016/j.apenergy.2017.06.079.

[104]  J. Ahn and S. Cho, "Development of an intelligent building controller to mitigate indoor thermal dissatisfaction and peak energy demands in a district heating system," *Building and Environment*, vol. 124, pp. 57–68, Nov. 2017, doi: 10.1016/j.buildenv.2017.07.040.

[105]  R. de Dear, G. Brager, and C. D., *Developing an Adaptive Model of Thermal Comfort and Preference - Final Report on RP-884.*, vol. 104. 1997.

[106]  A. Rogers, S. Ghosh, R. Wilcock, and N. R. Jennings, "A scalable low-cost solution to provide personalised home heating advice to households," in *Proceedings of the 5th ACM Workshop on Embedded Systems For Energy-Efficient Buildings*, 2013, pp. 1–8.

[107]  F. Auffenberg, S. Snow, S. Stein, and A. Rogers, "A Comfort-Based Approach to Smart Heating and Air Conditioning," *ACM Trans. Intell. Syst. Technol.*, vol. 9, no. 3, pp. 1–20, Dec. 2017, doi: 10.1145/3057730.

[108]  "Weather Data | EnergyPlus," *Weather data sources for energyplus framework*, 2015. https://energyplus.net/weather (accessed Mar. 14, 2020).

[109]  P. Danassis, K. Siozios, C. Korkas, D. Soudris, and E. Kosmatopoulos, "A low-complexity control mechanism targeting smart thermostats," *Energy and Buildings*, vol. 139, pp. 340–350, Mar. 2017, doi: 10.1016/j.enbuild.2017.01.013.

[110]  A. Javed, H. Larijani, A. Ahmadinia, R. Emmanuel, M. Mannion, and D. Gibson, "Design and Implementation of a Cloud Enabled Random Neural Network-Based Decentralized Smart Controller With Intelligent Sensor Nodes for HVAC," *IEEE Internet of Things Journal*, vol. 4, no. 2, pp. 393–403, Apr. 2017, doi: 10.1109/JIOT.2016.2627403.

[111]  A. Javed, H. Larijani, A. Ahmadinia, and D. Gibson, "Smart Random Neural Network Controller for HVAC Using Cloud Computing Technology," *IEEE Transactions on Industrial Informatics*, vol. 13, no. 1, pp. 351–360, Feb. 2017, doi: 10.1109/TII.2016.2597746.

[112]  A. Standard, "Standard 55-2010, Thermal environmental conditions for human occupancy," *American Society of Heating, Refrigerating and Air Conditioning Engineers*, 2010.

[113]  L. Jiang, R. Yao, K. Liu, and R. McCrindle, "An Epistemic-Deontic-Axiologic (EDA) agent-based energy management system in office buildings," *Applied Energy*, vol. 205, pp. 440–452, Nov. 2017, doi: 10.1016/j.apenergy.2017.07.081.

[114]  M. Deru *et al.*, "US Department of Energy commercial reference building models of the national building stock," 2011.

[115]  K. Konis and M. Annavaram, "The Occupant Mobile Gateway: A participatory sensing and machine-learning approach for occupant-aware energy management," *Building and Environment*, vol. 118, pp. 1–13, Jun. 2017, doi: 10.1016/j.buildenv.2017.03.025.

[116]  M. Macarulla, M. Casals, N. Forcada, and M. Gangolells, "Implementation of predictive control in a commercial building energy management system using neural networks," *Energy and Buildings*, vol. 151, pp. 511–519, Sep. 2017, doi: 10.1016/j.enbuild.2017.06.027.

[117]  D. Manjarres, A. Mera, E. Perea, A. Lejarazu, and S. Gil-Lopez, "An energy-efficient predictive control for HVAC systems applied to tertiary buildings based on regression techniques," *Energy and Buildings*, vol. 152, pp. 409–417, Oct. 2017, doi: 10.1016/j.enbuild.2017.07.056.

[118]  J. Reynolds, J.-L. Hippolyte, and Y. Rezgui, "A smart heating set point scheduler using an artificial neural network and genetic algorithm," in *2017 International Conference on Engineering, Technology and Innovation (ICE/ITMC)*, Funchal, Jun. 2017, pp. 704–710, doi: 10.1109/ICE.2017.8279954.

[119]  "Solar Resource Data and Tools | Grid Modernization | NREL." https://www.nrel.gov/grid/solar-resource/renewable-resource-data.html (accessed Mar. 15, 2020).

[120] "Southern Califirnia Edison," *SCE – Document Library*, 2017. https://library.sce.com/ (accessed Mar. 15, 2020).
[121] T. Wei, Y. Wang, and Q. Zhu, "Deep Reinforcement Learning for Building HVAC Control," in *Proceedings of the 54th Annual Design Automation Conference 2017 on - DAC '17*, Austin, TX, USA, 2017, pp. 1–6, doi: 10.1145/3061639.3062224.
[122] Yuan Wang, Kirubakaran Velswamy, and Biao Huang, "A Long-Short Term Memory Recurrent Neural Network Based Reinforcement Learning Controller for Office Heating Ventilation and Air Conditioning Systems," *Processes*, vol. 5, no. 4, p. 46, Aug. 2017, doi: 10.3390/pr5030046.
[123] B. Yuce and Y. Rezgui, "An ANN-GA Semantic Rule-Based System to Reduce the Gap Between Predicted and Actual Energy Consumption in Buildings," *IEEE Transactions on Automation Science and Engineering*, vol. 14, no. 3, pp. 1351–1363, Jul. 2017, doi: 10.1109/TASE.2015.2490141.
[124] D. Zhai, Y. C. Soh, and W. Cai, "Operating points as communication bridge between energy evaluation with air temperature and velocity based on extreme learning machine (ELM) models," in *2016 IEEE 11th Conference on Industrial Electronics and Applications (ICIEA)*, 2016, pp. 712–716.
[125] D. Zhai and Y. C. Soh, "Balancing indoor thermal comfort and energy consumption of ACMV systems via sparse swarm algorithms in optimizations," *Energy and Buildings*, vol. 149, pp. 1–15, Aug. 2017, doi: 10.1016/j.enbuild.2017.05.019.
[126] C. Zhong and J.-H. Choi, "Development of a Data-Driven Approach for Human-Based Environmental Control," *Procedia Engineering*, vol. 205, pp. 1665–1671, 2017, doi: 10.1016/j.proeng.2017.10.341.
[127] P. Carreira, A. A. Costa, V. Mansur, and A. Arsénio, "Can HVAC really learn from users? A simulation-based study on the effectiveness of voting for comfort and energy use optimization," *Sustainable Cities and Society*, vol. 41, pp. 275–285, Aug. 2018, doi: 10.1016/j.scs.2018.05.043.
[128] "EnergyPlus," *Energy.gov*. https://www.energy.gov/eere/buildings/energyplus-0 (accessed Mar. 15, 2020).
[129] D. 08-205-2015, "Design Standard for Energy Efficiency of Residential Buildings," http://www.shwjmc.cn/upfile/2018227158120.pdf, Shanghai, China, 2016.
[130] S. Gou, V. M. Nik, J.-L. Scartezzini, Q. Zhao, and Z. Li, "Passive design optimization of newly-built residential buildings in Shanghai for improving indoor thermal comfort while reducing building energy demand," *Energy and Buildings*, vol. 169, pp. 484–506, Jun. 2018, doi: 10.1016/j.enbuild.2017.09.095.
[131] L. A. Hurtado, E. Mocanu, P. H. Nguyen, M. Gibescu, and R. I. G. Kamphuis, "Enabling Cooperative Behavior for Building Demand Response Based on Extended Joint Action Learning," *IEEE Transactions on Industrial Informatics*, vol. 14, no. 1, pp. 127–136, Jan. 2018, doi: 10.1109/TII.2017.2753408.
[132] C. Marantos, C. P. Lamprakos, V. Tsoutsouras, K. Siozios, and D. Soudris, "Towards plug&play smart thermostats inspired by reinforcement learning," in *Proceedings of the Workshop on INTelligent Embedded Systems Architectures and Applications - INTESA '18*, Turin, Italy, 2018, pp. 39–44, doi: 10.1145/3285017.3285024.
[133] "Green Mark for Non-Residential Buildings NRB:2015," BCA Green Mark, Singapore, 2015. [Online]. Available: https://policy.asiapacificenergy.org/sites/default/files/Green Mark for Non-Residential Buildings NRB 2015.pdf.
[134] Y. Peng, A. Rysanek, Z. Nagy, and A. Schlüter, "Using machine learning techniques for occupancy-prediction-based cooling control in office buildings," *Applied Energy*, vol. 211, pp. 1343–1358, Feb. 2018, doi: 10.1016/j.apenergy.2017.12.002.
[135] A. Rajith, S. Soki, and M. Hiroshi, "Real-time optimized HVAC control system on top of an IoT framework," in *2018 Third International Conference on Fog and Mobile Edge Computing (FMEC)*, Barcelona, Apr. 2018, pp. 181–186, doi: 10.1109/FMEC.2018.8364062.
[136] P. H. Shaikh, N. B. M. Nor, P. Nallagownden, and I. Elamvazuthi, "Intelligent multi-objective optimization for building energy and comfort management," *Journal of King Saud University - Engineering Sciences*, vol. 30, no. 2, pp. 195–204, Apr. 2018, doi: 10.1016/j.jksues.2016.03.001.
[137] L. Zampetti, M. Arnesano, and G. M. Revel, "Experimental testing of a system for the energy-efficient sub-zonal heating management in indoor environments based on PMV," *Energy and Buildings*, vol. 166, pp. 229–238, May 2018, doi: 10.1016/j.enbuild.2018.02.019.
[138] Z. Zhang, A. Chong, Y. Pan, C. Zhang, S. Lu, and K. P. Lam, "A deep reinforcement learning approach to using whole building energy model for hvac optimal control," in *2018 Building Performance Analysis Conference and SimBuild*, 2018, p. 9.
[139] T. Chaudhuri, Y. C. Soh, H. Li, and L. Xie, "A feedforward neural network based indoor-climate control framework for thermal comfort and energy saving in buildings," *Applied Energy*, vol. 248, pp. 44–53, Aug. 2019, doi: 10.1016/j.apenergy.2019.04.065.
[140] "Pacific Gas & Electric - Tariffs," 2012. https://www.pge.com/tariffs/electric.shtml (accessed Mar. 16, 2020).
[141] Y. Chen, V. Chandna, Y. Huang, M. J. E. Alam, O. Ahmed, and L. Smith, "Coordination of Behind-the-Meter Energy Storage and Building Loads: Optimization with Deep Learning Model," in *Proceedings of the Tenth ACM*

*International Conference on Future Energy Systems - e-Energy '19*, Phoenix, AZ, USA, 2019, pp. 492–499, doi: 10.1145/3307772.3331025.

[142] T. N. S.L, "Climate Singapore / Changi Airport - Climate data (486980)," *www.tutiempo.net*. https://en.tutiempo.net/climate/ws-486980.html (accessed Jun. 07, 2020).

[143] G. Gao, J. Li, and Y. Wen, "Energy-Efficient Thermal Comfort Control in Smart Buildings via Deep Reinforcement Learning," *arXiv:1901.04693 [cs]*, Jan. 2019, Accessed: Mar. 09, 2019. [Online]. Available: http://arxiv.org/abs/1901.04693.

[144] "Weather Data by Region | EnergyPlus." https://energyplus.net/weather-region/asia_wmo_region_2/TWN%20%20 (accessed Mar. 16, 2020).

[145] W. Valladares *et al.*, "Energy optimization associated with thermal comfort and indoor air control via a deep reinforcement learning algorithm," *Building and Environment*, vol. 155, pp. 105–117, May 2019, doi: 10.1016/j.buildenv.2019.03.038.

[146] N. R. Canada, "canmeteo," Aug. 15, 2017. https://www.nrcan.gc.ca/maps-tools-publications/tools/data-analysis-software-modelling/canmeteo/19908 (accessed Apr. 06, 2020).

[147] N. Cotrufo, E. Saloux, J. M. Hardy, J. A. Candanedo, and R. Platon, "A practical artificial intelligence-based approach for predictive control in commercial and institutional buildings," *Energy and Buildings*, vol. 206, p. 109563, Jan. 2020, doi: 10.1016/j.enbuild.2019.109563.

[148] A. Ghofrani, S. D. Nazemi, and M. A. Jafari, "Prediction of building indoor temperature response in variable air volume systems," *Journal of Building Performance Simulation*, vol. 13, no. 1, pp. 34–47, Jan. 2020, doi: 10.1080/19401493.2019.1688393.

[149] A. H. Hosseinloo, A. Ryzhov, A. Bischi, H. Ouerdane, K. Turitsyn, and M. A. Dahleh, "Data-driven control of micro-climate in buildings; an event-triggered reinforcement learning approach," *arXiv:2001.10505 [cs, eess]*, Jan. 2020, Accessed: Apr. 04, 2020. [Online]. Available: http://arxiv.org/abs/2001.10505.

[150] A. Jain, F. Smarra, E. Reticcioli, A. D'Innocenzo, and M. Morari, "NeurOpt: Neural network based optimization for building energy management and climate control," *arXiv:2001.07831 [cs, eess]*, Jan. 2020, Accessed: Apr. 04, 2020. [Online]. Available: http://arxiv.org/abs/2001.07831.

[151] V. Limbachiya, K. Vadodaria, D. L. Loveday, and V. Haines, "Identifying a suitable method for studying thermal comfort in people's homes," in *Proceedings of the 7th Windsor Conference (Network for Comfort and Energy Use in Buildings) - The Changing Context of Comfort in an Unpredictable World*, Windsor, UK, Apr. 2012, p. 15 pp., [Online]. Available: https://repository.lboro.ac.uk/articles/Identifying_a_suitable_method_for_studying_thermal_comfort_in_people_s_homes/9339332.

[152] S. Javaid and N. Javaid, "Comfort evaluation of seasonally and daily used residential load in smart buildings for hottest areas via predictive mean vote method," *Sustainable Computing: Informatics and Systems*, vol. 25, p. 100369, Mar. 2020, doi: 10.1016/j.suscom.2019.100369.

[153] R. Lou, K. P. Hallinan, K. Huang, and T. Reissman, "Smart Wifi Thermostat-Enabled Thermal Comfort Control in Residences," *Sustainability*, vol. 12, no. 5, p. 1919, Mar. 2020, doi: 10.3390/su12051919.

[154] J. W. Moon and J. Ahn, "Improving sustainability of ever-changing building spaces affected by users' fickle taste: A focus on human comfort and energy use," *Energy and Buildings*, vol. 208, p. 109662, Feb. 2020, doi: 10.1016/j.enbuild.2019.109662.

[155] K. Moustris, K. A. Kavadias, D. Zafirakis, and J. K. Kaldellis, "Medium, short and very short-term prognosis of load demand for the Greek Island of Tilos using artificial neural networks and human thermal comfort-discomfort biometeorological data," *Renewable Energy*, vol. 147, pp. 100–109, Mar. 2020, doi: 10.1016/j.renene.2019.08.126.

[156] A. Sadeghi, R. Younes Sinaki, W. A. Young, and G. R. Weckman, "An Intelligent Model to Predict Energy Performances of Residential Buildings Based on Deep Neural Networks," *Energies*, vol. 13, no. 3, p. 571, Jan. 2020, doi: 10.3390/en13030571.

[157] C. Shan, J. Hu, J. Wu, A. Zhang, G. Ding, and L. X. Xu, "Towards non-intrusive and high accuracy prediction of personal thermal comfort using a few sensitive physiological parameters," *Energy and Buildings*, vol. 207, p. 109594, Jan. 2020, doi: 10.1016/j.enbuild.2019.109594.

[158] "INNOVA Thermal Comfort Booklet," [Online]. Available: http://www.labeee.ufsc.br/antigo/arquivos/publicacoes/Thermal_Booklet.pdf.

[159] L.-Y. Sung and J. Ahn, "Comparative Analyses of Energy Efficiency between on-Demand and Predictive Controls for Buildings' Indoor Thermal Environment," *Energies*, vol. 13, no. 5, p. 1089, Mar. 2020, doi: 10.3390/en13051089.

[160] R. Wang, S. Lu, and W. Feng, "A three-stage optimization methodology for envelope design of passive house considering energy demand, thermal comfort and cost," *Energy*, vol. 192, p. 116723, Feb. 2020, doi: 10.1016/j.energy.2019.116723.

[161] H. Xu, Y. He, X. Sun, J. He, and Q. Xu, "Prediction of thermal energy inside smart homes using IoT and classifier ensemble techniques," *Computer Communications*, vol. 151, pp. 581–589, Feb. 2020, doi: 10.1016/j.comcom.2019.12.020.